\documentstyle[epsf,emulateapj,apjfonts]{article}

\lefthead{SARAZIN, ET AL.}
\righthead{X-RAY PROPERTIES OF A1030/B2~1028+313}
\slugcomment{Astrophysical Journal, in press}

\begin{document}

\title{X-ray Properties of B2~1028+313:
A Quasar at the Center of the Abell Cluster A1030}

\author{Craig L. Sarazin}
\affil{Department of Astronomy, University of Virginia, 
P.O. Box 3818, Charlottesville, VA 22903-0818;
cls7i@virginia.edu,}

\author{Anton M. Koekemoer, Stefi A. Baum, Christopher P. O'Dea}
\affil{Space Telescope Science Institute, 3700 San Martin Drive, Baltimore,
MD 21218; sbaum@stsci.edu, koekemoe@stsci.edu, odea@stsci.edu,}

\author{Frazer N. Owen}
\affil{National Radio Astronomy Observatory, P.O Box 0, Socorro, NM 87801;
fowen@aoc.nrao.edu,}

\and

\author{Michael W. Wise}
\affil{Center for Space Research, Massachusetts Institute of Technology,
Bldg.\ 37-644, Cambridge, MA 02139; wise@space.mit.edu}

\begin{abstract}
X-ray observations with the {\it ROSAT} HRI and with {\it ASCA}
are presented for the nearby radio quasar B2 1028+313, which is located in
the cD galaxy at the center of the Abell cluster A1030.
We also analyze archival {\it ROSAT} PSPC observations.
We find that the X-ray emission is dominated by the quasar.
The flux varied by a factor of about two between the {\it ROSAT} HRI
and {\it ASCA} observations, which were about one year apart.
The X-ray spectrum of the quasar is fit by a single power-law, except
at low energies where there is a soft excess.
Although the shape of the soft excess is not strongly constrained, it
can be fit by a blackbody with a temperature of about 30 eV.
There was evidence for extended X-ray emission, which contributed about
25\% of the total flux.
However, this emission does not appear to be normal X-ray emission from
intracluster gas or a central cooling flow.
The extended X-ray emission appears to be quite soft;
if its spectrum is modeled as thermal emission, the temperature is
$\sim$0.2 keV, rather than the 5-10 keV expected for ICM emission.
The radial surface distribution of the emission was not fit
by either the beta model which usually describes ICM emission, or by a
cooling flow model.
The {\it ASCA} and {\it ROSAT} spectra showed no convincing evidence for a
thermal component with a cluster-like temperature, either in the overall
spectral shape or in emission lines.
In addition, the {\it ROSAT} PSPC image showed that the extended X-ray
emission was highly elongated to the NNW and SSE, in the same direction
as the extended radio emission from the quasar.
We suggest that the extended emission is inverse Compton emission
from the extended radio lobes.
\end{abstract}

\keywords{
galaxies: clusters: general ---
galaxies: clusters: individual (A1030) ---
galaxies: cD ---
intergalactic medium ---
quasars: individual (B2~1028+313) ---
X-rays: galaxies
}

\section{Introduction} \label{sec:intro}
B2~1028+313 is a well-known, nearby ($z = 0.1782$) radio-loud quasar,
At radio wavelengths the quasar has a bright, inverted-spectrum core
and a one-sided radio jet to the NNW
(Gower \& Hutchings 1984).
On larger scales, the radio emission extends to the NNW and SSE
(Owen, White, \& Ge 1993;
Owen \& Ledlow 1997).
The radio axis varies with radius around a mean position angle
of $-30^\circ$ (measured from the north to the east).
The quasar is a bright X-ray source
(Zamorani et al.\ 1981;
Blumenthal, Keel, \& Miller 1982;
Wilkes \& Elvis 1987),
whose spectrum shows evidence for a soft X-ray excess above
a power-law spectrum
(Elvis, Wilkes, \& Tannabaum 1985;
Wilkes \& Elvis 1987;
Masnou et al.\ 1992;
Zhou \& Yu 1992).
The optical spectrum of the quasar is rich in emission lines,
and shows a strong rise to the blue
(Jackson \& Browne 1991;
Shastri et al.\ 1993;
Owen, Ledlow, \& Keel 1995, 1996).
The quasar is also quite bright in the UV
(Lanzetta, Turnshek, \& Sandoval 1993).

Recently, Owen, Ledlow, \& Keel (1995) noticed that the quasar is located
in the central cD galaxy of A1030, an Abell richness class 0 cluster
(Abell, Corwin, \& Olowin 1989).
At low redshift, it is uncommon for quasars to be located in the center
of a cluster, although higher redshift radio quasars are more strongly
clustered
(e.g., Yee \& Ellingson 1993).
The host galaxy of B2~1028+313 is a large, very bright elliptical galaxy which
is similar to other cD galaxies in its optical properties
(Ledlow \& Owen 1995).
The major axis of the galaxy is at a position angle of 45$^\circ$,
which is about 75$^\circ$ away from the radio axis.

The location of this nearby radio quasar in the center of a rich cluster
suggests that some portion of the X-ray emission from the system
may be coming from intracluster gas.
The X-ray luminosity determined from the $Einstein$ Observatory data
was $L_X = 7.1 \times 10^{44}$ erg s$^{-1}$ (0.3--3.5 keV:
Wilkes \& Elvis 1987), which is typical for rich Abell clusters.
On the other hand, A1030 is a apparently a poor cluster.
The X-ray flux is about 30 times higher than would be expected
based on the correlation between core X-ray and radio fluxes of
core-dominated radio quasars
(Worrall et al.\ 1994).
The core in B2~1028+313 provides more than 50\% of the 5 GHz radio flux
(Gower \& Hutchings 1984;
Condon et al.\ 1994), so that B2~1028+313 is core-dominated.
On the other hand, its ratio of X-ray to optical flux is more consistent
with values for lobe-dominated radio quasars
(Worrall et al.\ 1994), where the X-ray emission may have a large
contribution from thermal emission by ambient gas.

The presence of this bright, nearby quasar in the center of a cluster
provides an important opportunity to study the interaction between
the AGN and its environment.
If a portion of the X-ray emission is due to the presence of a hot intracluster
medium (ICM), then it is likely that the radio source may interact with this
thermal plasma.
The radio source associated with the quasar is quite compact
(Owen \& Ledlow 1997), as might be expected if it were confined by
dense ICM.
Moreover, the quasar is surrounded by a luminous emission-line region
extended on scales $\sim$25 kpc
(Owen et al.\ 1998),
typical of those found in cooling-flow clusters
(e.g., Heckman et al.\ 1989).
Thus, it is possible that the ICM in the region around the quasar is
particularly dense, because of a cooling flow.

The location of this quasar in the center of a cluster also provides a
light source for searching for absorption by cooler material in the
ICM.
In this regard, the fact that B2~1028+313 is particularly bright
in the UV is very useful.
Recently, we used {\it HST} UV spectra to set very strong limits
on the absorbing column of any cooler material in the cluster
(Koekemoer et al.\ 1998).
One might also be able to detect cool or hot components of the
ICM as absorption lines or edges in the X-ray spectrum of the quasar
(Sarazin 1989;
Wise \& Sarazin 1997).

In this paper, we present {\it ROSAT} PSPC and HRI images and
{\it ASCA} and {\it ROSAT} PSPC spectra of the X-ray emission
from B2~1028+313 and the surrounding A1030 cluster.
The first object of these observations are to determine the X-ray properties
of the quasar.
Secondly, we want to detect or limit the X-ray emission from intracluster
gas in A1030, and to search for evidence for a central cooling flow in this
cluster.
Third, we want to study the interaction of the radio source with
its environment, and
search for X-ray emission associated with the extended
radio lobes.
Finally, if there is evidence for dense ICM around the quasar, we will
search for X-ray absorption features in the quasar spectrum due to the
ambient gas.

The X-ray observations are presented in \S~\ref{sec:data}.
The observed X-ray fluxes are compared and variability detected in
\S~\ref{sec:flux}.
In \S~\ref{sec:spatial}, we search for extended X-ray emission in the
cluster, present X-ray images, and determine the radial surface
brightness profile.
The X-ray spectra from {\it ROSAT} and {\it ASCA} are presented
in \S~\ref{sec:spectra}, and are compared to combinations of nonthermal
models for emission by the AGN and thermal models for ICM emission.
We also place limits on any emission or absorption lines in
\S~\ref{sec:spectra_lines}.
The results are discussed in \S~\ref{sec:discussion}, where
we review the X-ray properties of the quasar, and consider the limits
on the thermal emission by the ICM.
The relationship between the extended X-ray and radio emission
is discussed in \S~\ref{sec:discussion_radio}.
All distance-dependent values in this paper assume
$H_o = 50$ km s$^{-1}$ Mpc$^{-1}$ and $q_o = 0.5$.
Unless otherwise stated, all of the uncertainties are at the
90\% confidence level.

\section{X-ray Observations} \label{sec:data}

\subsection{{\it ROSAT} HRI Observations} \label{sec:data_hri}

A1030 was observed with the {\it ROSAT} High Resolution Imager (HRI)
during the period 11-27 May, 1996.
The total exposure time was 18,600 seconds.
In addition to the normal processing of the data, we examined the
light curve of a large source-free region to check for any periods
of enhanced background, and none were found.  
Because we are interested in achieving the best possible angular
resolution, we also examined the aspect history for any anomalies during
the accepted time in the image.
None were found.
The X-ray image was corrected for particle background, exposure, and
vignetting using the SXRB software package of Snowden (1995).
In addition to A1030/B2~1028+313, 14 other rather weak X-ray sources
appear in the HRI image.
Only one of these could be identified, and it corresponded to an
otherwise unidentified X-ray source from the $Einstein$ catalog,
2E~1027.6+3112.

\subsection{{\it ROSAT} PSPC Observations} \label{sec:data_pspc}

A1030 was also observed with the {\it ROSAT} Position Sensitive Proportional
Counter (PSPC).
We have extracted these observations from the public archive to help 
constrain the spatial extent and spectral properties of the emission
from this system.
A1030 was observed for 2,481 seconds on 5, February 1992
(RP700433; P.I.\ B. Wilkes).
(Actually, the observation was much longer than this, but most
of the data were unusable, apparently because of problems with the
aspect solution.)
A longer observation of 6,906 seconds was made on the same day
with the boron filter in place.
The extra opacity of this filter means that the unfiltered and
filtered observations produced a similar number of photons from A1030.
The PSPC data were screened for periods of high background based on
a Master Veto Rate $>$ 170
(Plucinsky et al.\ 1993),
for other times of high background,
for periods of 15 seconds after switching to the high voltage,
and for periods with an uncertain aspect solution.
The resulting live exposure times for the two images were
2,155 seconds for the unfiltered image and 6698 seconds for
the image with the boron filter.
The average values of the Master Veto Rate were 79.6 for the unfiltered
data and 78.6 for the boron filter data.

\subsection{{\it ASCA} Observations} \label{sec:data_asca}

A1030 was observed with {\it ASCA} in two intervals on
1995 May 3-4 and May 7.
The two pointings were nearly identical, and have been combined in
all of the analysis.
Although the data were originally processed with Revision 1 of the
standard processing, they were screened using the standard Revision
2 screening criteria.
In addition, light curves of the data were constructed and searched for
otherwise unrejected data in periods with high background or data
dropouts.
The total exposures were approximately 36.8 ksec for the SIS
and 39.0 ksec for the GIS.

The Solid-state Imaging Spectrometer (SIS) detectors operated with only
one chip per detector being active (S0C1 and S1C3).
In order to maximize the spectral resolution and the accuracy of
the calibration, all of the SIS data were taken in Faint mode and analyzed 
in Bright2 mode.
The SIS data were corrected for the echo effect and dark frame error using
{\sc ftools} {\sc faint} version 3.11.
The SIS PI channels were determined using {\sc sispi} version 1.1 using
the gain file of 110397.
Hot and flickering pixels were removed with {\sc cleansis} version 1.6.
A level discriminator (at approximately 0.42 keV) was used to avoid telemetry
saturation when SIS data were taken at Medium data rate.

During the period of this observation, the GIS-3 spectral gain was sometimes
overestimated by the standard ASCA processing
(Idesawa et al.\ 1997).
This was corrected using the
{\sc ftools} {\sc temp2gain} version 4.1 with the 1996 May 12 version of
the gis\_temp2gain calibration file.

\section{X-ray Fluxes and Variability} \label{sec:flux}

In the {\it ROSAT} HRI observations, a total of $2763 \pm 61$ counts were
detected from A1030, after corrections for background, exposure, and vignetting.
This corresponds to a flux of $4.18 \times 10^{-12}$ ergs s$^{-1}$ in
the observed 0.2 -- 2.0 keV frame, uncorrected for absorption.
We assumed a power-law spectrum with a photon index of $\Gamma = 2.08$
and an absorbing column of $N_H = 1.21 \times 10^{20}$ cm$^{-2}$.

\begin{figure*}[tbh]
\plotone{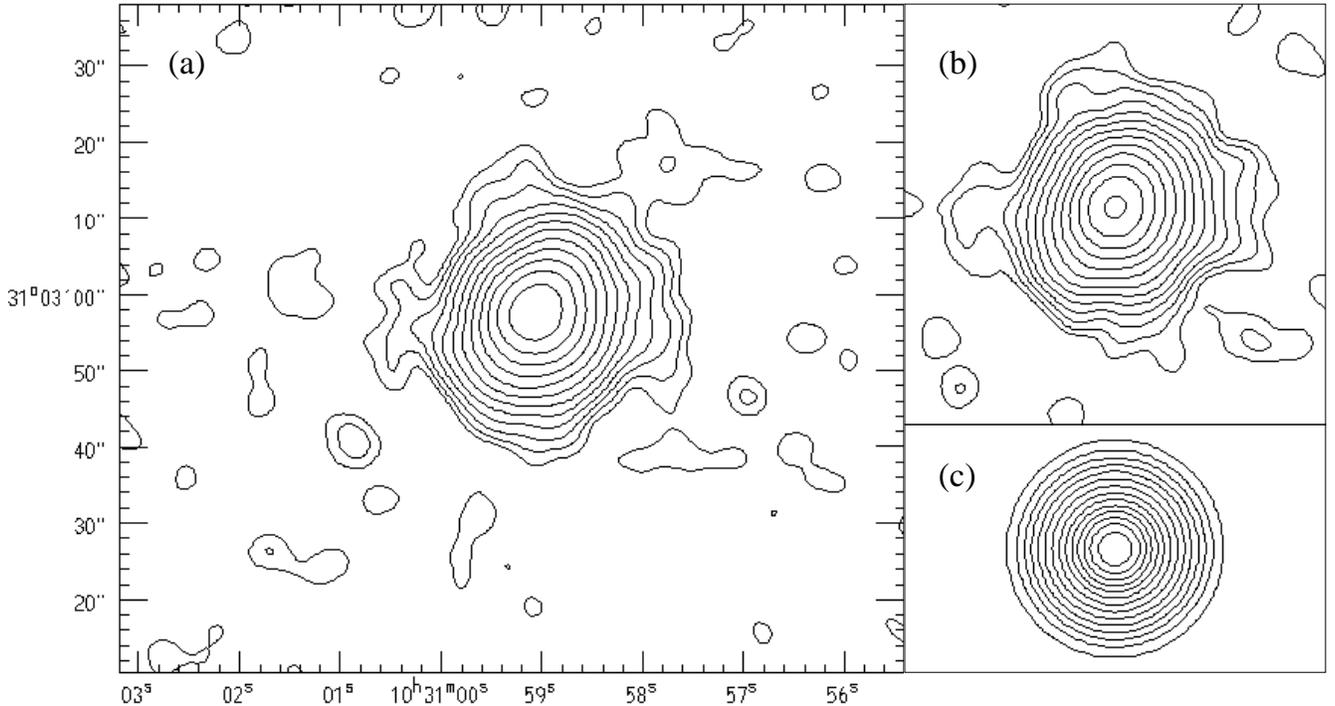}
\vskip-0.3truein
\caption{
{\it (a)} Contours of the central {\it ROSAT} HRI X-ray image of A1030.
The image was smoothed with a $\sigma = 2\arcsec$ gaussian.
The coordinates are J2000.
There are 14 contours logarithmicly spaced between 0.02 and 10
counts/pixel (1 pixel $= 0.5\arcsec \times 0.5\arcsec$).
{\it (b)} The center of the HRI image of A1030 after ``correcting''
for the effects of the wobble on the aspect solution, as discussed in
the text.
The angular scale and contours are the same as in {\it (a)}.
{\it (c)} A contour map of the nominal {\it ROSAT} HRI
Point Response Function (PRF).
The smoothing, scale, and contours are the same as in {\it (a)} and {\it
(b)}.
\label{fig:hri_contour}}
\end{figure*}

\centerline{\null}


%
%
\tabcaption{\hfil X-Ray Flux vs.\ Observation Date \label{tab:flux} \hfil}
\begin{center}
\begin{tabular}{lcc}
\tableline
\tableline
Instrument      &Observation    &$f_\nu$(1 keV)\cr
                &Date           &($\mu$Jy)\cr
\tableline
Einstein IPC    &1979 May 24\phs\phn\phn        &$0.74^{+0.09}_{-0.04}$\cr
ROSAT PSPC      &1992 Feb 5\phn\phn\phn\phn\phn &$0.75^{+0.06}_{-0.05}$\cr
ASCA GIS,SIS    &1995 May 3$-$7\phn\phn         &$0.43^{+0.02}_{-0.02}$\cr
ROSAT HRI       &1996 May 11$-$27               &$0.84^{+0.04}_{-0.03}$\cr
\tableline
\end{tabular}
\end{center}
\vskip0.2truein

The observed flux for the unfiltered PSPC observation was
$3.80 \times 10^{-12}$ ergs s$^{-1}$ (0.2 -- 2.0 keV),
uncorrected for absorption.
The flux in the same band using the boron filtered data was
$4.08 \times 10^{-12}$ ergs s$^{-1}$.
These values are within 10\% of the HRI flux, measured four years later.
The luminosity of A1030 in the rest frame {\it ROSAT} band 0.2--2 keV is
$L_X = 7 \times 10^{44}$ ergs s$^{-1}$, corrected for absorption.

In order to compare the fluxes measured at different epochs using different
instruments, we also determined the observed flux density $f_\nu$ at 1 keV,
uncorrected for absorption.
We adopted this energy because it was within the band pass of {\it ROSAT},
{\it Einstein}, and {\it ASCA}, and this value was derived by
Wilkes \& Elvis (1987) from the {\it Einstein} IPC observation.
In Table~\ref{tab:flux}, the flux densities measured by the different
instruments at different times are given in $\mu$Jy
(1 $\mu$Jy $\equiv 10^{-29}$ ergs cm$^{-2}$ s$^{-1}$ Hz$^{-1}$).
The uncertainties are at the 90\% confidence level for a single interesting
parameter.
Only statistical uncertainties are included;
the calibration errors are probably at least as large as the
statistical uncertainties in most cases.
The second column gives the dates of the observations.
The {\it ROSAT} PSPC and HRI fluxes assume the best-fit single
power-law fit to the {\it ROSAT} PSPC X-ray spectrum
(\S~\ref{sec:spectra_pspc}).
Also, the flux listed was derived from the unfiltered PSPC observation.
The observation with the boron filter, which occurred on the same date
as the unfiltered observation, gave a flux which was 7\% higher
(\S~\ref{sec:spectra_pspc}).
This is within the uncertainties in the fluxes;
it is also likely that the calibration errors in the two instruments are
of this order.
The {\it ASCA} flux is based on the SIS0, GIS2, and GIS3 instruments.
The flux in SIS1 was slightly lower (\S~\ref{sec:spectra_asca}), but
consistent within the likely calibration uncertainties.

The fluxes from the {\it Einstein} and {\it ROSAT} PSPC are remarkably
consistent, given the passage of 13 years.
The flux from the {\it ROSAT} HRI is only slightly higher, and marginally
consistent within the uncertainties.
On the other hand, the observed flux during the {\it ASCA} observation was
about a factor of two lower than any of the other fluxes.
This result does not depend on the particular spectral model used to
convert count rates into fluxes for each of the instruments.
Thus, we conclude that although the X-ray flux may have remained
approximately constant for much of the time since 1979, it must on
occasion vary by at least a factor of two on time scale as short as a year.
On the other hand, there was no significant evidence for any variation
during the period of {\it ROSAT} HRI observation, during the
{\it ROSAT} PSPC observation, or during the {\it ASCA} observation.
This may indicate that the source doesn't vary by a large factor
on time scales as short as a few days to two weeks.

The variability of the source on time scales of a year means that
at least 50\% of the X-ray emission is from the quasar rather than
the cluster.

\section{Spatial Structure} \label{sec:spatial}

\subsection{{\it ROSAT} HRI Spatial Distribution} \label{sec:spatial_hri}

Figure~\ref{fig:hri_contour}a shows the contour map of the inner
approximately $100\arcsec \times 90\arcsec$ of the HRI image of A1030.
The image was smoothed with a gaussian kernel with $\sigma = 2\arcsec$,
and the 14 contours are logarithmically spaced between 0.02 and 10
counts/pixel (1 pixel $= 0.5\arcsec \times 0.5\arcsec$).
The centroid of the X-ray emission was at
R.A.~= 10$^{\rm h}$30$^{\rm m}$59\fs1 and
Dec.~= 31$^\circ$02\arcmin58\arcsec (J2000),
which is within 3$\arcsec$ of the position of the quasar B2~1028+313.
For comparison,
Figure~\ref{fig:hri_contour}c shows a contour map of the nominal
{\it ROSAT} HRI Point Response Function (PRF) on the axis of
the instrument (David et al.\ 1993), smoothed in the same way as the
data.
The image of A1030 appears somewhat more extended than that expected
from a point source, and is elongated at a position angle ($PA$) of about
-29$^\circ$ (measured counterclockwise from north).
None of the other sources in the HRI image are bright enough to
usefully test for a similar extension in the image.

The HRI observation of A1030 was done in 6 OBIs (observational
intervals) during the period 11-27 May, 1996.
The image in Figure~\ref{fig:hri_contour}a resulted from
combining all of these observations.
To check for possible offsets in the aspect solution between these
OBIs, we produced images from each of the separate OBIs.
They all gave a consistent centroid for the X-ray emission,
and all showed an elongation in the same general direction
($PA \approx -29^\circ$).

During most pointed observations with {\it ROSAT}, the satellite is
``wobbled'' by several arcmin with a period of several hundred seconds.
Morse (1994) has suggested that errors in the aspect solution during
wobbling can result in extended and elongated images of point sources.
Although he argues that the aspect errors are associated with the wobble,
Morse finds that the observed elongations of the images of point sources
aren't necessarily parallel to the projection of the wobble on the sky.
The Nominal Roll Angle for the observation of A1030 was $-14.98^\circ$
(measured clockwise from north).
This implies that the projection of the wobble on the sky
was at $PA \approx -30^\circ$.
Unfortunately, this is very close to the observed direction of elongation
of the central part of the image of A1030 in Figure~\ref{fig:hri_contour}a.
This increases the concern that extent and elongation of our HRI image
of A1030 might be due to the effects of the wobble, rather than due to
extended emission from the cluster.

Morse (1994) developed a technique to attempt to correct for aspect
errors due to the wobble.
Unfortunately, his original technique can only be applied to very bright
sources.
In Appendix~\ref{sec:aspect}, we present a variant on Morse's algorithm
which can be applied to fainter sources.
The resulting image of the center of A1030 is shown in
Figure~\ref{fig:hri_contour}b.
The center part of this image is considerably narrower than
the original image (Fig.~\ref{fig:hri_contour}a).
However, beyond a few arcsec, the ``wobble-corrected'' image is
very similar to the original image, and is elongated in nearly
the same direction ($PA = 23^\circ$).
The wobble-corrected image still appears somewhat more elongated than
the HRI PRF (Fig.~\ref{fig:hri_contour}c).
We conclude that the wobble-corrected image in Figure~\ref{fig:hri_contour}b
actually is a better representation of the correct image of the
central region of A1030.
However, the restriction to a finite set of regions on the detector
implies  that this image is missing flux at larger distances from the
center of the A1030, and we use the original image beyond the
central 10\arcsec, and for determining the total flux from
the source.

\centerline{\null}
\vskip3.6truein
\includegraphics{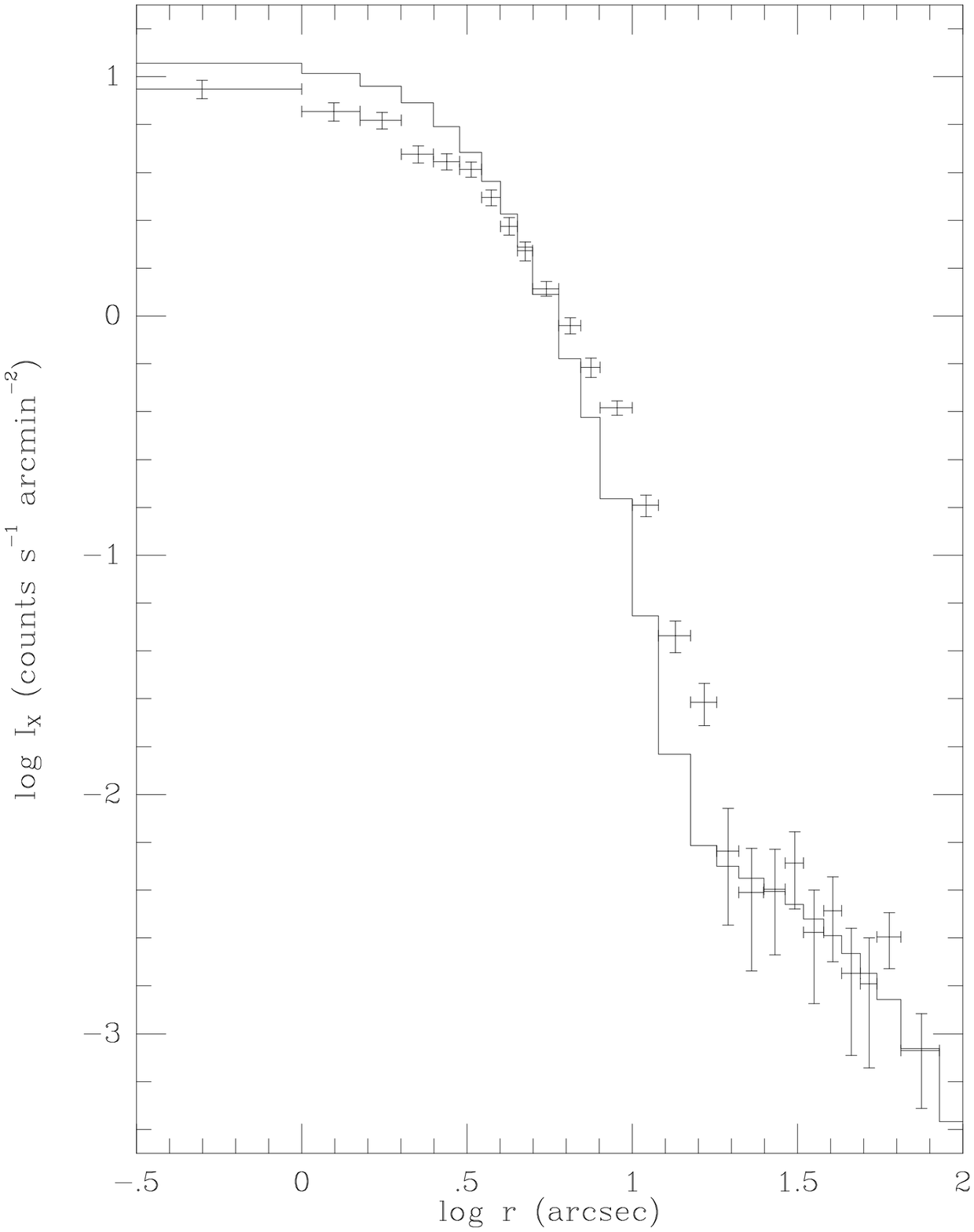}
\figcaption[fig2.ps]{
The {\it ROSAT} HRI surface brightness of A1030 as a function of the
radius, corrected for background, exposure, and vignetting.
The data points give the surface brightness measured in circular
annuli with 1-$\sigma$ uncertainties.
The solid histogram is the HRI Point Response Function
(PRF), normalized to the data and accumulated in the same annuli as
the data.
\label{fig:hri_surf}}
\vskip0.2truein

To determine or limit the flux from A1030 which is due to extended
cluster emission, we accumulated the HRI counts in annuli centered
on the X-ray centroid.
All of the annuli had at least 20 counts to assure the applicability
of Gaussian statistics.
Other X-ray sources were excluded.
The resulting surface brightness profile for the wobble-corrected image
is shown in Figure~\ref{fig:hri_surf}.
The data points give the observed surface brightness, corrected for
background, exposure, and vignetting.
For comparison, we constructed an image of the {\it ROSAT} PRF,
and normalized it to have the same number of counts within a
radius of 100\arcsec\ as the actual image.
Note that the broad wings on the image ($r \ga 15\arcsec$) are apparently
just due to the PRF;
this is the ``halo'' produced by the electrostatic screen on the
HRI.
This halo severely limits our ability to detect extended faint emission
with the HRI.
However, in the central regions, even the wobble-corrected image
is broader than the PRF.
This broadening is most evident at radii of
$r \approx 10\arcsec \approx 40$ kpc.
The best-fit point source model for the image has $\chi^2 = 264.6$ for
25 degrees of freedom (d.o.f.).


%
%
\begin{table*}[htb]
\caption{\hfil Fits to the ROSAT X-Ray Surface Brightness
\label{tab:spatial} \hfil}
\begin{center}
\begin{tabular}{llccccc}
\tableline
\tableline
Detector&Model&$r_X$&$\beta$&$\gamma$&Extended Fraction&$\chi^2$/d.o.f.\cr
&&(arcsec)&&&(\%)&\cr
\tableline
HRI &Point Source&                &             &               &
(0)             & 264.6/25 = 10.58\cr
HRI &Gaussian   &$5.2 \pm 0.5$    &             &               &
$57^{+6}_{-5}$ & 25.2/23 = 1.09\cr
HRI &Beta Model &$31^{+4}_{-20}$  &$>$2.02      &(0)            &
$42^{+3}_{-3}$ & 22.6/22 = 1.03\cr
HRI &Cooling Flow&$31^{+5}_{-20}$ &$>$2.02      &$<$0.91        &
$42^{+6}_{-5}$ & 22.6/21 = 1.08\cr
&&&&&&\cr
PSPC&Point Source&                &             &               &
(0)              & 218.4/25 = 8.74\cr
PSPC&Gaussian   &$32 \pm 4$       &             &               &
$35^{+2}_{-2}$ & 25.2/23 = 1.09\cr
PSPC&Beta Model &$55^{+143}_{-26}$&$>$0.95      &               &
$29^{+3}_{-2}$& 18.5/22 = 0.84\cr
PSPC &Cooling Flow&$55^{+219}_{-26}$&$>$0.95    &$<$1.41        &
$29^{+7}_{-5}$ & 18.5/21 = 0.88\cr
\tableline
\end{tabular}
\end{center}
\end{table*}

We tried a number of simple models to fit the HRI surface brightness.
Some of the results are shown in Table~\ref{tab:spatial}, which also gives
the results of fits to the PSPC surface brightness (\S~\ref{sec:spatial_hri}).
All of the models include a point source at the center.
The model surface brightness was convolved with the PRF of the instrument,
and accumulated in the same annuli used for the data.
The best fit model was determined by minimizing $\chi^2$.
The quoted uncertainties are 90\% confidence regions for a single interesting
parameter ($\Delta \chi^2 = 2.706$).
For each model, the Table gives a spatial scale parameter $r_X$,
and the fraction of the flux within 2\arcmin\ radius in the extended component.
The best-fit value of $\chi^2$ and the number of degrees of freedom
are also given.
Parameters in parentheses are fixed by assumption.

In general, we found that a large number of models could fit the
extended component of the HRI surface brightness, because it is only
seen clearly over a small range of radii.
Each of these fits required a point source as well as an extended
component.
For example, Table~\ref{tab:spatial} shows the results of fitting a
Gaussian surface brightness component (where $r_X = \sigma$ for
the Gaussian).
We also fit the standard isothermal beta model,
\begin{equation}
I_X (r) = I_o \left[ 1 + \left( \frac{r}{r_X} \right)^2
\right]^{-3 \beta + 1/2} \, .
\label{eq:beta_model}
\end{equation}
For the isothermal beta model, the value of $\beta$ is also given.
In the beta model, the value of $\beta$ is not very well constrained, but
the lower limit of 2.02 is much larger than the value found for typical
clusters ($\sim$0.6).
This implies that the X-ray surface brightness of the extended component
in A1030 falls off much faster at large radii than would be expected for
cluster emission.
Note that the value of the core radius $r_X \approx 31\arcsec \approx 120$
kpc, which is a reasonable value for the core radius of a cD cluster.

We also tried a model which provides a reasonable fit to the X-ray surface
brightness profiles of many nearby clusters with cooling flows.
It has the same form as the beta model at large radii, but the
surface brightness increases as a power-law function of the radius
at small radii:
\begin{equation}
I_X (r) = I_o
\left( \frac{r}{r_X} \right)^{-\gamma}
\left[ 1 + \left( \frac{r}{r_X} \right)^2 \right]^{-3 \beta + 1/2} \, .
\label{eq:cf_model}
\end{equation}
For nearby cooling flow clusters, $\gamma \approx 1$
(e.g., Fabian 1994).
The spatial scale $r_X$ in this model is more closely related
to the cooling radius in the cluster than to the core radius.
The resulting of fitting this model to the HRI observations are shown in
Table~\ref{tab:spatial}.
The best-fit value of $\gamma$ is 0.03, which is quite small.
This makes the cooling flow model nearly identical to the beta model.
Thus, there is no strong evidence for a cooling flow from the
{\it ROSAT} HRI image.

\subsection{{\it ROSAT} PSPC Spatial Distribution} \label{sec:spatial_pspc}

\begin{figure*}[tbh]
\vskip3.6truein
\includegraphics{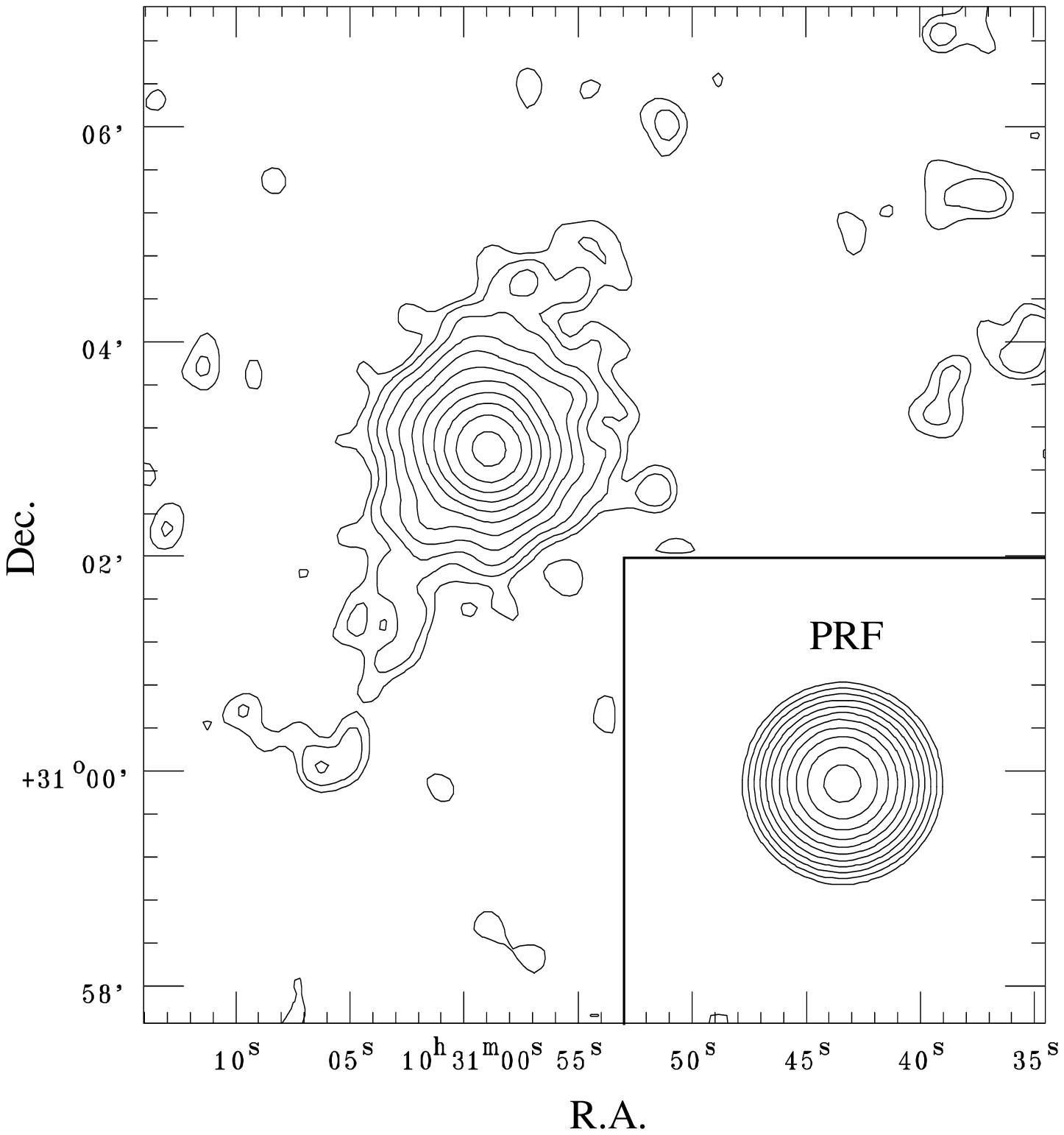}
\caption{
Contours of the central {\it ROSAT} PSPC X-ray image of A1030.
The image is based on the sum of the unfiltered and boron filter data.
The image has been corrected for background and vignetting, but not
for exposure.
The image was smoothed with a $\sigma = 8\arcsec$ gaussian.
The coordinates are J2000.
There are 11 contours logarithmicly spaced between 0.05 and 20
counts/pixel (1 pixel $= 4\arcsec \times 4\arcsec$).
The panel in the lower right corner gives the {\it ROSAT} PSPC PRF
for this image, smoothed and contoured in the same way as the data.
\label{fig:pspc_contour}}
\end{figure*}

We also studied the spatial extent of the X-ray emission in A1030
using the {\it ROSAT} PSPC data.
Although we initially attempted to determine the image and surface brightness
profile separately for the unfiltered and boron filter data, and to
use only the hard band (0.52--2.0 keV) data, there were too few counts.
Thus, we broadened the photon energy band used ($ 11 \le PI \le 235$),
and merged the unfiltered and boron filter data.
Figure~\ref{fig:pspc_contour} shows X-ray contours of the merged
{\it ROSAT} PSPC image the central region of A1030.
The image has been corrected for background and vignetting, but not
exposure as it is the sum of the unfiltered and boron filter image
with different exposures and responses.
The image was smoothed with a gaussian kernel with $\sigma = 8\arcsec$,
For comparison, the lower right panel shows the {\it ROSAT} PSPC PRF
for these data, after the same smoothing as the data.
The PRF of the PSPC is energy dependent.
To determine the correct PRF to apply to these data, we determined
the distribution of counts within 2\arcmin\ radius of the center as a
function of the photon energy channel ($PI$).
Then we constructed a number of PRFs for different photon energies,
and weighted them by the observed fraction of counts in the
relevant range of $PI$ values.

The observed image of A1030 is wider than the PRF at radii $\ga$25\arcsec.
While the morphology of the extension at small radii is unclear from
the PSPC image, at radii beyond $\sim$1\arcmin\ the X-ray emission is
strongly elongated to the NNW and SSE at a position angle of about
$-30^\circ$.
This is similar to the direction of elongation of the HRI image
(Figure~\ref{fig:hri_contour}).
In addition to the elongation, two compact X-ray sources are
located at nearly equal distances of about 3\farcm5 from the nucleus of
B2~1028+313 along approximately the same position angle as the X-ray elongation
(at
10$^h$30$^m$51$^s$, 31$^\circ$06\arcmin,
and
10$^h$31$^m$06$^s$, 31$^\circ$00\arcmin).
These sources also appear in the {\it ROSAT} HRI image.
It is unclear whether they are extended or point sources, or whether they
are related to B2~1028+313 or not.

The extended X-ray emission in Figure~\ref{fig:pspc_contour} is very soft.
Although there are too few photons to extract a spectrum, we determined
X-ray colors for the inner 1\arcmin\ circle and 1\arcmin-3\arcmin\ annulus
about the center.
First, we determined an X-ray color from
the ratio of the counts in the Snowden (1995)
R4-R7 bands (approximately 0.52--2.02 keV) to those in the Snowden
R1L-R2 bands (approximately 0.11--0.41 keV) in the unfiltered PSPC image.
In the inner 1\arcmin, this ratio is $0.91 \pm 0.10$.
There is no significant detection of hard photons in the
1\arcmin--3\arcmin\ annulus, and the
upper limit of the color is $<$0.29 at the 90\% confidence level.
This radial variation in the color is too large to be due to the energy
dependent PRF of the {\it ROSAT} PSPC.
Second, we determine a X-ray color from the ratio of the boron
filtered PSPC count rate to that for the unfiltered PSPC image
for the 1\arcmin--3\arcmin\ annulus.
This gave a ratio of $0.21 \pm 0.09$.
Both of these values of the X--ray color are quite soft.
If the emission were thermal from diffuse gas with a
heavy element abundance of 0.3 of solar and subject to Galactic absorption,
the temperature of the gas would be $kT \approx 0.20$ keV.
Alternatively, if it is due to a power-law component with Galactic
absorption, the photon index of the power law is $\Gamma = 2.9 \pm 0.4$.
With this spectrum, the count rates imply a total flux density
at 0.2 keV of about 0.7 $\mu$Jy in the 1\arcmin--3\arcmin\ annulus.
If we adopt the same power-law index as observed for the radio emission
from the same region ($\alpha = 1.45$, $\Gamma = 2.45$,
\S~\ref{sec:discussion_radio}), then the flux at 0.2 keV is about
0.5 $\mu$Jy.

We also determined the radial surface brightness profile of A1030 from
the PSPC data.
Counts were accumulated in annuli centered on the X-ray centroid.
The observed surface brightness was corrected for background, exposure,
and vignetting.
Other X-ray sources were excluded.
We separately extracted the surface brightness profiles for the
unfiltered and boron filter data, and corrected them for
background, relative exposure, and vignetting.
The two profiles agreed to within the uncertainties.
Then, we added the two profiles together to improve the signal-to-noise
ratio.
All of the annuli had at least 20 counts to assure the applicability
of Gaussian statistics.
The resulting merged PSPC surface brightness profile 
is shown in Figure~\ref{fig:pspc_surf}.
Note that we have not normalized the surface brightness to the
total exposure, as this is not really meaningful for the sum of
the unfiltered and boron filtered data.

The instrument PRF discussed above was normalized to have the same number
of counts within a radius of 2\arcmin\ as the actual image, and
was accumulated in the same annuli as the data.
The resulting PRF surface brightness is shown as a solid histogram in
Figure~\ref{fig:pspc_surf}.
The best-fit point source model for the surface brightness has
$\chi^2 = 218.4$ for 25 d.o.f.

\centerline{\null}
\vskip3.7truein
\includegraphics{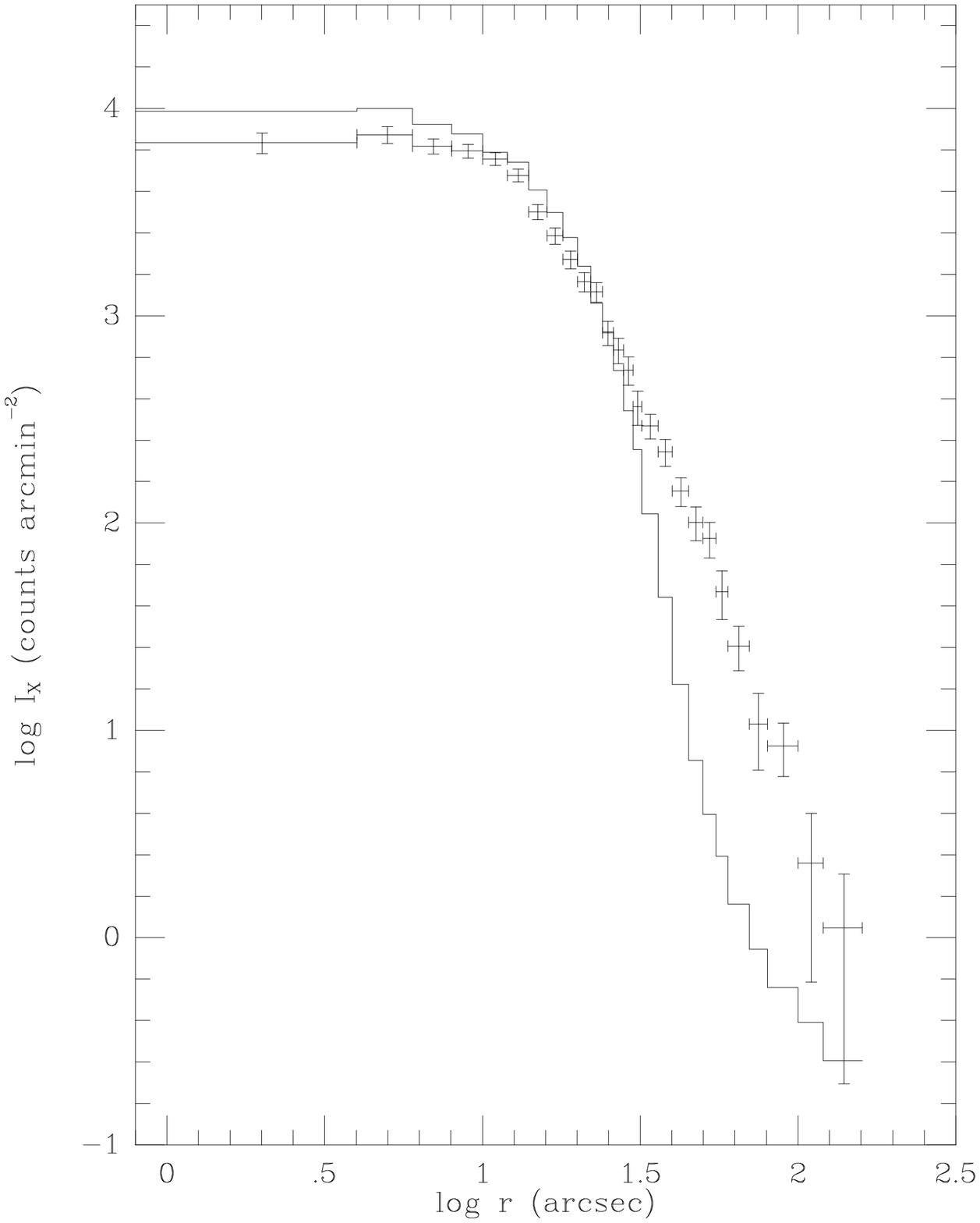}
\figcaption[fig4.ps]{
The {\it ROSAT} PSPC surface brightness of A1030 as a function of the
radius, corrected for background, exposure, and vignetting.
The data points give the surface brightness measured in circular
annuli with 1-$\sigma$ uncertainties.
The data are the sum of the surface brightness in the unfiltered
and boron filtered data.
The solid histogram is the PSPC Point Response Function
(PRF), normalized to the data and accumulated in the same annuli as
the data.
\label{fig:pspc_surf}}
\vskip0.2truein

We also included an extended component in the fits, which was modeled
using a Gaussian, the beta model (eq.~\ref{eq:beta_model}), or
the cooling flow model (eq.~\ref{eq:cf_model}).
The results are shown in Table~\ref{tab:spatial}.
The beta model provided an acceptable fit, but the value of
$\beta$ was larger than found for typical nearby clusters.
The addition of a cooling flow at the center did not improve
the fit significantly.

Given that the extended X-ray emission in the PSPC image is highly
elongated, has a very soft spectrum, and does not follow the surface
brightness profile expected for the ICM, it seems unlikely that
the X-rays in the elongated features in Figure~\ref{fig:pspc_contour}
are due to normal intracluster gas.
To limit the X-ray contribution from normal ICM, we determined the
X-ray emission in Figure~\ref{fig:pspc_contour} in an annulus between
1\arcmin\ and 3\arcmin, but excluding wedges which contained the
X-ray extensions to the NNW and SSE.
Only $49 \pm 20$ counts were detected above the background in this
region, of which $7 \pm 6$ were expected from spill over of the
PRF from the central source.
Thus, the net count was $42 \pm 21$.
For comparison, the total count above background within 3\arcmin
was $2276 \pm 81$.
Correcting for the area of the X-ray elongation which were excluded 
from this determination, we find that any circularly symmetric ICM emission
at projected radii beyond 1\arcmin\ accounts for $<$4\% of the X-ray
emission in the {\it ROSAT} PSPC band.

\section{X-ray Spectra} \label{sec:spectra}

\subsection{{\it ROSAT} PSPC Spectra} \label{sec:spectra_pspc}

The PSPC spectra from the unfiltered and boron filter observations
were extracted from a circle of 2\arcmin\ radius from the center of
A1030.
The spectra were corrected for background. 
The photon energy channels below 0.2 keV and above 2.2 keV
were discarded, and the spectra were grouped so that each
spectral bin had at least 20 counts.
The spectra were fit using the XSPEC program.
Initially, the Galactic absorbing column was fixed at
$N_H = 1.98 \times 10^{20}$ cm$^{-2}$.
Based on NRAO Greenbank 140 ft.\ observations towards this target,
this should be a reasonably accurate value
(Elvis, Lockman, \& Wilkes 1989).
We fit the emission from the source as a power law, with the
number of emitted photons per unit photon energy $E$ varying as
$E^{-\Gamma}$.
The unfiltered and boron filter data were initially fit separately;
the results are shown in rows 1 and 2 of Table~\ref{tab:spectra}.
The two fits gave values of the photon spectral index $\Gamma$
in reasonably agreement.
The boron filter data suggested a slightly higher normalization
(by about 5\%) than the unfiltered data.
This difference would seem to be within the range of calibration
uncertainties.


%
%
\begin{table*}[htb]
\caption{\hfil ASCA and ROSAT X-Ray Spectral Fits \label{tab:spectra} \hfil}
\begin{center}
\begin{tabular}{llcccccc}
\tableline
\tableline
Row&Instrument&$N_H$&Power Law&\multicolumn{3}{c}{Blackbody or Thermal
Model}&
$\chi^2$/d.o.f.\cr
\cline{5-7}
&&&$\Gamma$&$kT$&Abund.&Fraction&\cr
&&($10^{20}$ cm$^{-2}$)&&(keV)&(solar)&(\%)&\cr
\tableline
 1&PSPC Unfilt.&(1.98)                &$2.32^{+0.08}_{-0.09}$&
&&(0)&41.3/36 = 1.15\cr
 2&PSPC Boron  &(1.98)                &$2.47^{+0.13}_{-0.15}$&
&&(0)&41.2/42 = 0.98\cr
 3&PSPC        &(1.98)                &$2.35^{+0.08}_{-0.07}$&
&&(0)&84.7/79 = 1.07\cr
 4&PSPC        &1.21$^{+0.41}_{-0.37}$&$2.08 \pm 0.17$       &
&&(0)&75.9/78 = 0.97\cr
 5&PSPC        &(1.98)                &$2.16 \pm 0.15$       &
$0.022^{+0.013}_{-0.015}$&BB&$9^{+1}_{-4}$ & 74.8/77 = 0.97\cr
 6&PSPC        &(1.98)                &                      &
0.82&0.0&(100)&165.8/78 = 2.13\cr
&&&&&&&\cr
 7&ASCA GIS    &(1.98)                &$1.71 \pm 0.06$       &
&&(0)&231.2/245 = 0.94\cr
 8&ASCA GIS    &$<$5.91               &$1.70 \pm 0.07$       &
&&(0)&231.1/244 = 0.95\cr
 9&ASCA SIS    &(1.98)                &$1.84 \pm 0.04$       &
&&(0)&258.5/287 = 0.90\cr
10&ASCA SIS    &$5.62^{+2.45}_{-2.35}$&$1.95 \pm 0.08$       &
&&(0)&252.0/286 = 0.88\cr
11&ASCA        &(1.98)                &$1.80 \pm 0.03$       &
&&(0)&498.9/534 = 0.93\cr
12&ASCA        &$2.71^{+1.99}_{-1.92}$&$1.82 \pm 0.06$       &
&&(0)&498.5/533 = 0.94\cr
13&ASCA        &(1.98)                &                      &
$5.34^{+0.40}_{-0.38}$&$0.22^{+0.11}_{-0.11}$&(100)&562.4/533 = 1.06\cr
14&ASCA        &(1.98)                &$1.80^{+0.06}_{-0.04}$&
(7)&(0.3)&$<$21&498.9/533 = 0.94\cr
&&&&&&&\cr
15&ASCA \& PSPC&(1.98)                &$1.88 \pm 0.03$       &
&&(0)&693.0/614 = 1.13\cr
16&ASCA \& PSPC&$0.61^{+0.16}_{-0.15}$&$1.78 \pm 0.03$       &
&&(0)&587.4/613 = 0.96\cr
17&ASCA \& PSPC&(1.98)                &$1.81 \pm 0.03$       &
$0.028^{+0.012}_{-0.055}$&BB&$15 \pm 1$&586.9/612 = 0.96\cr
18&ASCA \& PSPC&$0.56^{+0.15}_{-0.15}$&$1.80^{+0.12}_{-0.05}$&
(7)&(0.3)&$<$21&589.2/614 = 0.96\cr
\tableline
\end{tabular}
\end{center}
\end{table*}

\begin{figure*}[htb]
\vskip4.0truein
\includegraphics{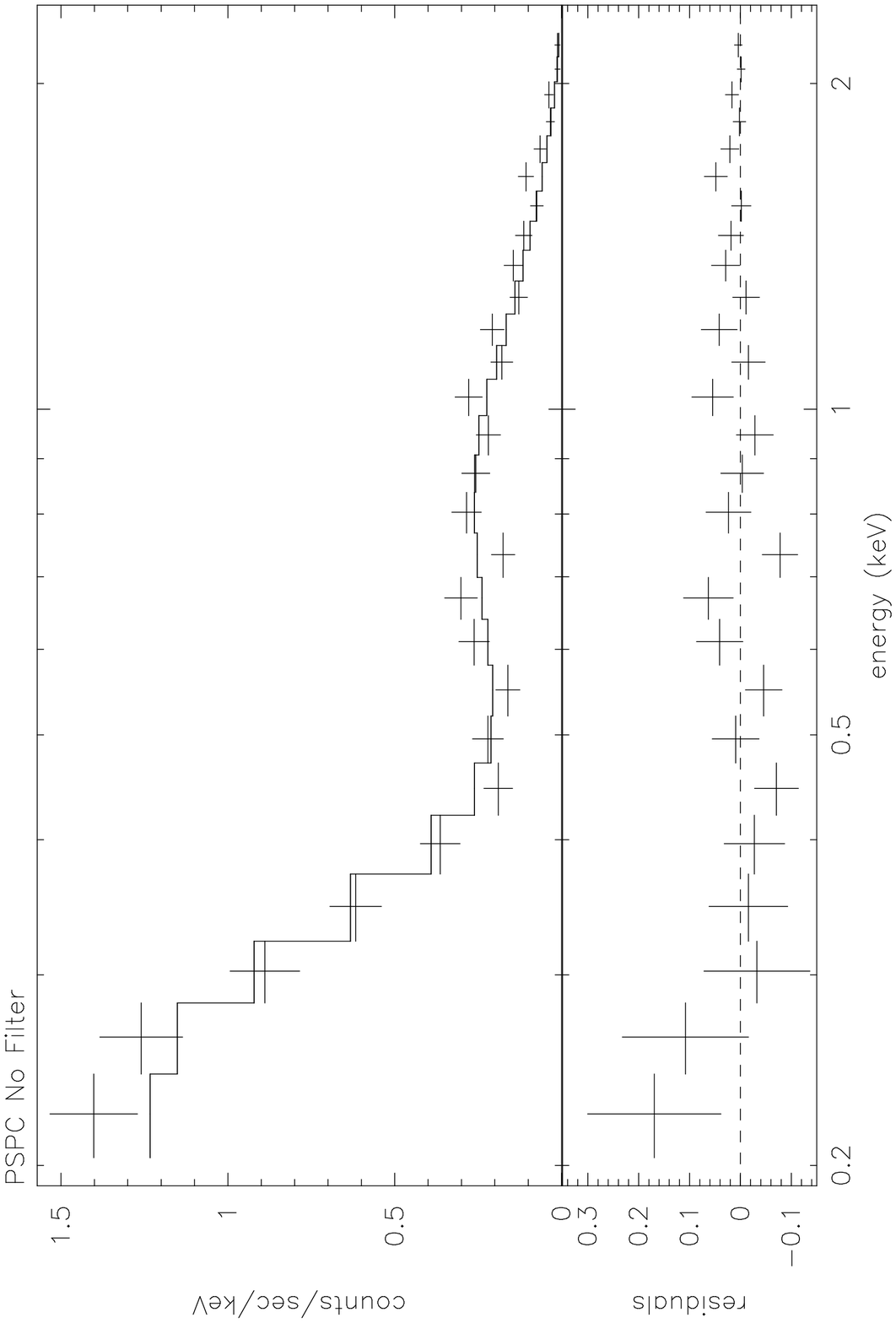}
\caption{
The {\it ROSAT} X-ray spectrum of A1030 for the unfiltered data.
The upper panel gives the data and the best-fit power-law model
for the unfiltered and boron filter data
(row 3 of Table~\protect\ref{tab:spectra}) assuming the measured
Galactic absorption of $N_H = 1.98 \times 10^{20}$ cm$^{-2}$,
The crosses give the data points with 1-$\sigma$ uncertainties bars,
while the histogram is the model.
The width of the data points or histogram steps is the width of
the energy channels used to accumulate the data.
The lower panel gives the residuals to the fit (in counts/sec/keV).
\label{fig:pspc_spect}}
\end{figure*}

\begin{figure*}[htb]
\vskip4.2truein
\includegraphics{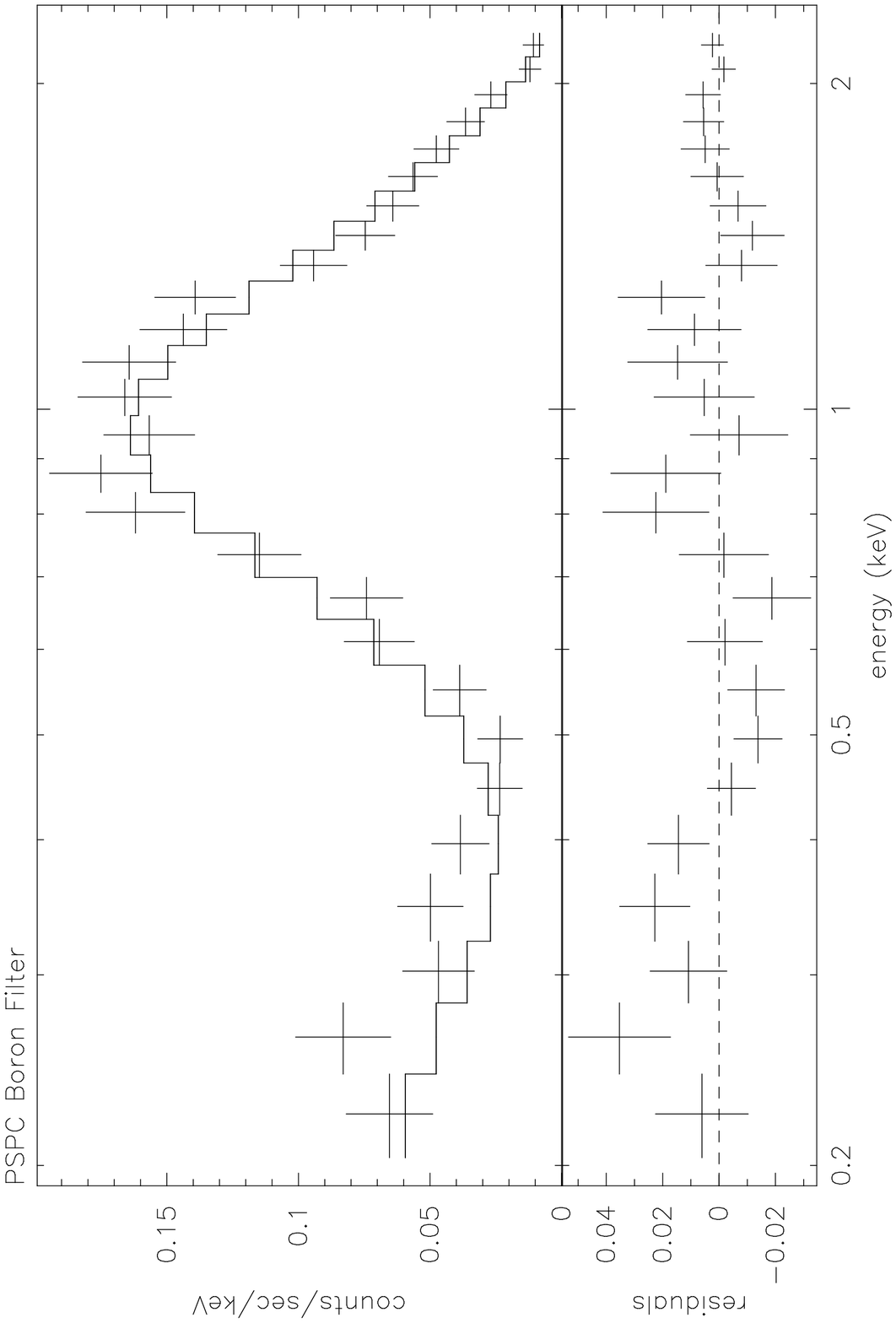}
\caption{
The {\it ROSAT} X-ray spectrum of A1030 for the boron filter data.
The notation and model are the same as in
Figure~\protect\ref{fig:pspc_spect}.
\label{fig:pspc_spect_boron}}
\end{figure*}

We then fit the two PSPC spectra jointly to a single power-law, while
allowing the normalization of the spectra to differ slightly.
The result is shown in row 3 of Table~\ref{tab:spectra}.
Again, this is a reasonably good fit.
These fits to the spectra are shown in Figures~\ref{fig:pspc_spect}
and \ref{fig:pspc_spect_boron}.

There is some evidence for a soft X-ray excess in the data from both of the
instruments.
A similar result was found by Wilkes \& Elvis (1987) based on
the $Einstein$ IPC spectrum of this object.
As one method of fitting this soft excess, we allowed the Galactic absorbing
column to vary.
This significantly improved the fit to the PSPC spectra (row 4
of Table~\ref{tab:spectra}).
For comparison, Wilkes \& Elvis (1987) found that the $Einstein$ IPC
spectrum was fit by a power-law with $\Gamma = 1.5^{+0.3}_{-0.1}$
and $N_H \le 1.3 \times 10^{20}$ cm$^{-2}$.
The {\it ROSAT} PSPC spectral index is somewhat larger than the
{\it Einstein} value, but this may reflect the inadequacy of modeling
the soft X-ray as reduced X-ray absorption, and the wider energy band pass
of {\it Einstein}.

It is more physically plausible to model the soft X-ray excess as an
additional emission component.
None of the spectra presented in this paper strongly constrain the shape
of this component.
As a result, we have chosen arbitrarily to model the soft X-ray excess
as a blackbody.
Row~5 in Table~\ref{tab:spectra} shows the result of a fit, fixing the
absorbing column at the Galactic value, but adding a soft blackbody
component.
The temperature is not very strongly constrained.
This excess soft X-ray component provides about 9\% of the
total 0.2--10 keV X-ray luminosity of the quasar
(the fraction listed in Table~\ref{tab:spectra}).

We also tried to fit the PSPC spectrum using a single temperature
thermal {\sc mekal} model for emission by the intracluster medium.
A purely thermal model for the PSPC X-ray spectrum did not provide
an acceptable fit, and required very low temperatures and abundances
for the intracluster gas
(row 6 of Table~\protect\ref{tab:spectra}), which are not consistent
with the values found in other clusters.

\subsection{{\it ASCA} Spectra} \label{sec:spectra_asca}

The GIS and SIS spectra of the source were extracted using a circular
regions with radii of 6\arcmin\ and 4\arcmin, respectively.
GIS background spectra were extracted using the same regions of the
detectors from the 1994 May blank sky calibration fields.
The GIS background data were cleaned in the same way as the source data.
The histogram of the values of the Cut Off Rigidity ($COR$) during the
observation was determined, and the blank sky background field observations
were summed so as to reproduce the $COR$ distribution of the A1030
observation for both GIS2 and GIS3.
We also tried using backgrounds accumulated from symmetric positions
on the detectors during our own observation.
These gave consistent results, although the statistic uncertainties were much
smaller for the blank sky backgrounds.

Determining an appropriate background spectrum for the SIS observations
proved to be more difficult.
We tried accumulating local backgrounds from our own observation.
However, in 1-CCD mode, there is very little area left over on the chip
once the source region is excluded.
Moreover, the background spectrum is then accumulated from a less sensitive
region of the detector than the source spectrum.
The blank sky SIS observations in the calibration database were all taken
in 4-CCD bright mode, and so have a somewhat different response than the
data in our observation.
Also, the calibration observations were done early in the mission, and
the SIS detectors have degraded significantly since then.
We found that these observations gave background spectra which were
inconsistent with our observations.
Instead, we used a blank sky observation done at about the same time as
our observation and extracted from the archive (observation 73016000).
This high Galactic latitude observation was done in 1-CCD faint mode,
but without the use of the level discriminator.
The background spectra were accumulated from the same regions of the
detectors as our source spectra.

The GIS and SIS spectra were limited to the photon energy ranges of
0.55--11 keV and 0.45--11 keV, respectively.
The spectra were binned so that each channel contained at least
20 counts, so that $\chi^2$ statistics could be applied.
The two GIS instruments were found to give consistent spectral fits,
both in terms of the shape of the spectra and the normalization.
The normalization of the SIS0 spectrum was also consistent with that of
the two GIS detectors.
However, while the two SIS instruments gave consistent fits for the
shape of the spectrum, the SIS1 spectra give slightly lower fluxes (by about
5\%).
Thus, the normalization of the SIS1 instrument was allowed to vary
relative to the others.
All of the spectral fitting was done with separate response matrixes for
all four {\it ASCA} instruments.
However, for the purpose of plotting the spectra and models, we
created total GIS and SIS spectra, averaging over the responses of
each of the two detectors.
Also, in the plots of the overall spectra, we summed up to five adjacent
channels to make the plots less confusing.

\begin{figure*}[htb]
\vskip4.4truein
\includegraphics{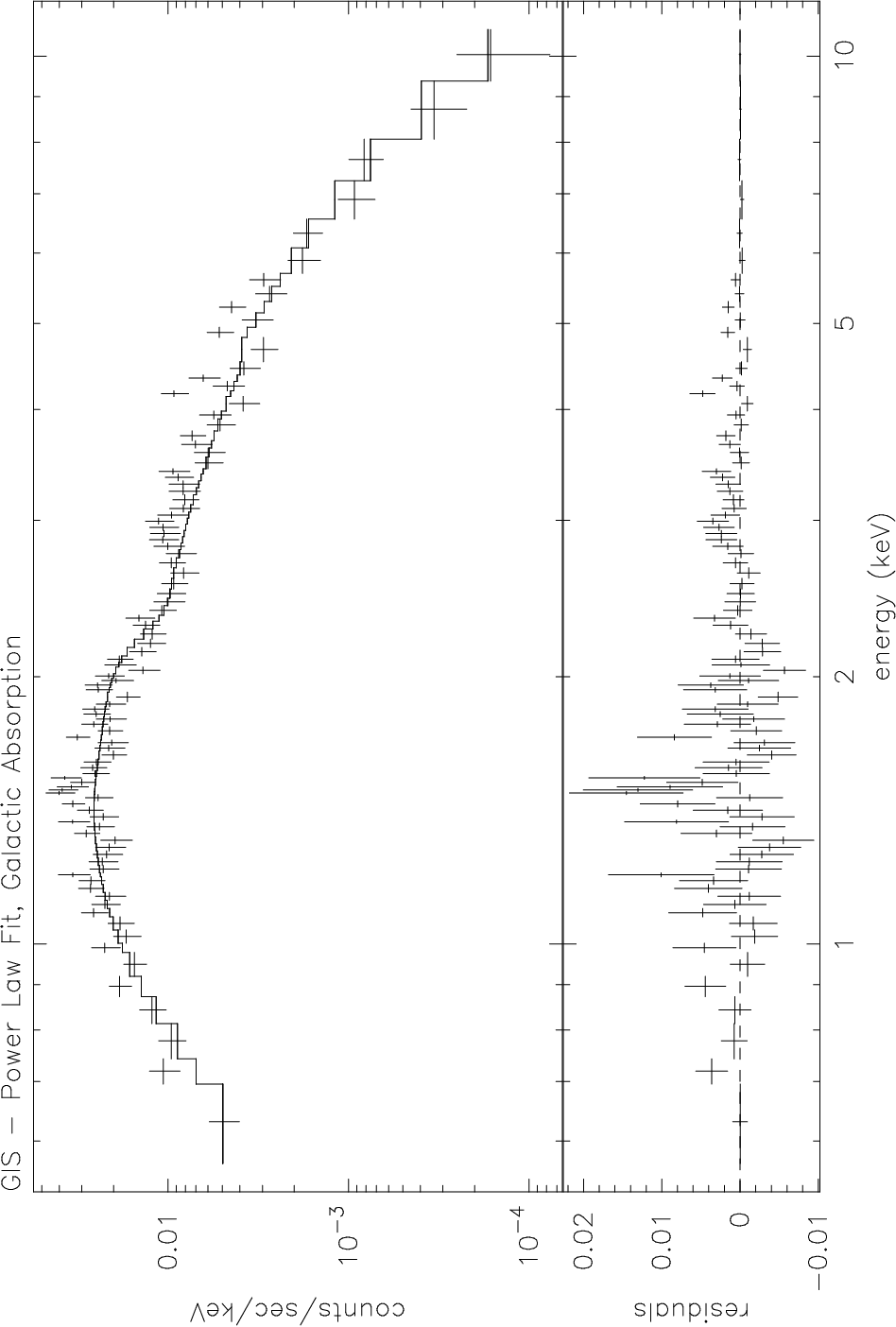}
\caption{
The {\it ASCA} GIS X-ray spectrum of A1030.
The upper panel gives the data and the best-fit power-law model
(row~7 of Table~\protect\ref{tab:spectra}) assuming the measured
Galactic absorption of $N_H = 1.98 \times 10^{20}$ cm$^{-2}$,
The notation is the same as in
Figure~\protect\ref{fig:pspc_spect}.
\label{fig:gis_spect}}
\end{figure*}

The best-fit power-law spectral fit to the GIS2 and GIS3 spectra assuming
the measured Galactic column are shown in row~7 of Table~\ref{tab:spectra}.
A power-law is a reasonably good fit to the GIS spectra, although
the slope is considerably flatter than that required to fit the PSPC
spectra.
The total GIS spectrum and best-fit model are shown in
Figure~\ref{fig:gis_spect}.
The GIS has too little soft X-ray sensitivity to strongly constrain the
absorbing column;
if the absorption is allowed to vary, only an upper limit is found
(row~8 of Table~\ref{tab:spectra}).

\begin{figure*}[htb]
\vskip4truein
\includegraphics{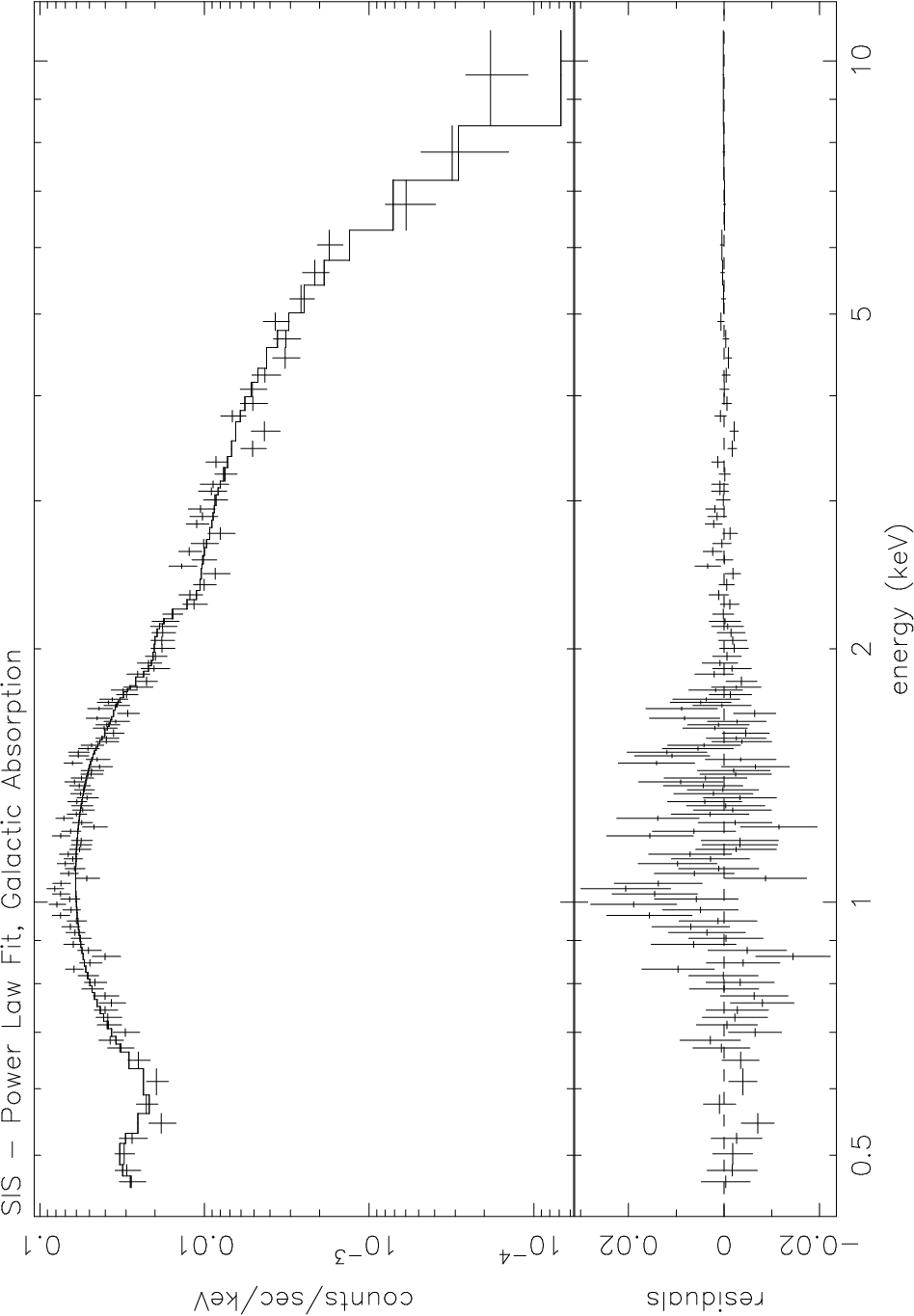}
\caption{
The {\it ASCA} SIS X-ray spectrum of A1030.
The upper panel gives the data and the best-fit power-law model
(row~10 of Table~\protect\ref{tab:spectra}) assuming the measured
Galactic absorption of $N_H = 1.98 \times 10^{20}$ cm$^{-2}$,
The notation is the same as in
Figure~\protect\ref{fig:pspc_spect}.
\label{fig:sis_spect}}
\end{figure*}

A power-law spectrum also was a good fit to the SIS data.
The best-fit model assuming the Galactic column is shown in row~9 of
Table~\ref{tab:spectra} and Figure~\ref{fig:sis_spect}.
The power-law slope is intermediate between that derived from
the GIS and that from the PSPC.
The observed spectrum is below the model in the region of the
oxygen K-edge; when the absorption is allowed to vary, the
best-fit absorption value is considerably above the Galactic
value (row~10 of Table~\ref{tab:spectra}).

We also fit a power-law model jointly to the {\it ASCA} GIS and SIS
spectra.
The results for Galactic or variable absorption are shown in rows
11 and 12 of Table~\ref{tab:spectra}.
A single power-law spectrum with Galactic absorption provided an acceptable
fit to the SIS and GIS spectra;
allowing the absorption to vary did not improve the fit significantly.
The best-fit joint spectrum is somewhat inconsistent with the individual
SIS and GIS spectral fits; for Galactic absorption, these gave
$\chi^2 = 489.7$ for 532 d.o.f.
This difference is statistically significant;
however, the difference could also be explained by a small ($\sim$3\%)
systematic error in the responses of the GIS and/or SIS.

A purely thermal model for the {\it ASCA} spectrum provided an adequate fit
(row~13 of Table~\ref{tab:spectra}).
Unlike the thermal fit to the PSPC spectrum (row~6 of Table~\ref{tab:spectra}),
the temperature and abundance were not unreasonable.
However, this is a much poorer fit than the power-law model.
We also tried fitting the {\it ASCA} spectra assuming both a power-law
and thermal contribution.
The best-fit thermal contribution was zero.
Obviously, the parameters of the thermal model could not be determined,
so we assumed $k T = 7$ keV and an abundance of 0.3 of solar.
As shown in row 14 of Table~\ref{tab:spectra}, the maximum thermal
contribution to the spectrum was 21\% of the 0.2--10 keV rest frame
X-ray luminosity,
or $L_X (thermal) < 1.6 \times 10^{44}$ ergs s$^{-1}$.

\subsection{Joint Fits to {\it ROSAT} and {\it ASCA} Spectra}
\label{sec:spectra_both}

We also tried to fit simultaneously the {\it ROSAT} PSPC unfiltered
and boron filter spectra along with the {\it ASCA} SIS and GIS spectra.
Of course, between the {\it ROSAT} PSPC observation and the {\it ASCA}
observation, the flux from the source dropped by a factor of nearly
two (Table~\ref{tab:flux}).
Thus, we are assuming that the shape of the spectrum remained unchanged
while the flux varied, which may very well be wrong.
Obviously, we have to allow the normalization of the spectrum to differ
between the {\it ROSAT} PSPC and {\it ASCA} spectra.
Row~15 in Table~\ref{tab:spectra} shows the best joint spectra fit,
assuming a power-law spectrum and the Galactic absorbing column.
This is not a terrible fit, but it is much worse than the separate
fits to the {\it ROSAT} PSPC and {\it ASCA} spectra.
This may indicate that the spectrum varied somewhat between this two
epochs.

Alternatively, the {\it ROSAT} PSPC spectra suggested the presence of
an excess soft X-ray component.
If we model this by allowing the absorbing column to vary, the
joint {\it ASCA} and PSPC spectral fit is considerably improved
(row~16 of Table~\ref{tab:spectra}).
It is not quite as good a fit as the fits to the individual spectra,
and the difference is statistically significant.
On the other hand, the difference could also be accounted for by
a small systematic error of $\sim$3\% in the responses of the instruments.
If we model the soft X-ray excess as a blackbody, the spectral parameters
are shown in row 17 of Table~\ref{tab:spectra}.
The temperature of the blackbody is poorly determined but is about
30 eV.
This soft component provides about 15\% of the 0.2--10 keV luminosity
of the quasar.

In order to provide another limit on the X-ray contribution of intracluster
gas, we fit the joint {\it ROSAT} PSPC and {\it ASCA} spectra to a model
with both a power-law component and a thermal component.
Since the X-ray emission of the intracluster gas cannot vary on observable
time scales, we required that the normalization of the thermal component
be constant.
It would also be difficult to understand how the absorbing column toward
the intracluster gas could be lower than the Galactic value, so we fixed
this quantity as well.
The best-fit spectral model had only a small thermal contribution, and the
spectral properties of this component were not well-determined.
For the purposes of limiting the thermal contribution, we fixed the
gas temperature at $kT = 7$ keV and the abundances at 0.3 of solar.
The results are shown in row~18 of Table~\ref{tab:spectra}.
The absorbing column value applies to the power-law component.
The upper limit on the contribution of the intracluster gas is
21\% of the 0.2--10 keV rest frame X-ray luminosity during the {\it ROSAT}
PSPC observation (the value in Table~\ref{tab:spectra}), or 36\% of the
{\it ASCA} luminosity.
The upper limit on the intracluster gas luminosity is
$L_X (thermal) < 2.6 \times 10^{44}$ ergs s$^{-1}$
(0.2--10 keV).

We also tried modeling the cluster emission assuming a cooling flow
spectrum.
The upper limits here are even tighter: $<$8\% of the {\it ROSAT} flux
or $<$14\% of the {\it ASCA} flux, corresponding to limits on the
luminosity of $L_X (thermal) < 0.9 \times 10^{44}$ ergs s$^{-1}$
(0.5--10 keV) or limits on the cooling rate of
$\dot{M} < 93 \, M_\odot$ yr$^{-1}$.

%
%
\tabcaption{\hfil ASCA Limits on Emission and Absorption Lines
\label{tab:lines} \hfil}
\scriptsize
\begin{center}
\begin{tabular}{lccc}
\tableline
\tableline
Line&Redshifted Line Energy&\multicolumn{2}{c}{Equivalent Width Limits
(eV)}\cr
\cline{3-4}
&(keV)&Emission&Absorption\cr
\tableline
Fe {\sc XXVI} 1s--2p       & 5.91 & \phn41 &    111 \cr
Fe {\sc XXV} 1s$^2$--1s2p  & 5.67 & \phn44 &    106 \cr
S {\sc XVI} 1s--2p         & 2.23 & \phn43 & \phn79 \cr
S {\sc XV} 1s$^2$--1s2p    & 2.09 & \phn95 & \phn68 \cr
Si {\sc XIV} 1s--2p        & 1.70 &    125 & \phn79 \cr
Si {\sc XIII} 1s$^2$--1s2p & 1.58 & \phn33 & \phn54 \cr
\tableline
\end{tabular}
\end{center}
\normalsize
\vskip0.2truein

\subsection{Limits on Emission or Absorption Lines}
\label{sec:spectra_lines}

In addition to its overall effect on the shape of the X-ray spectrum,
any thermal intracluster gas component would produce X-ray line
emission.
We have searched for evidence of line emission in the {\it ASCA}
X-ray spectra of A1030 without success.
For this purpose, we fit the X-ray spectrum in the region around
relatively isolated, strong lines with a power-law continuum plus
a Gaussian line.
The redshifted line energy and width were determined from the
various line contributions to each line feature.
The strength of the line was allowed to vary.
The fits were done using unbinned spectra and the C-statistic.
For the Fe K lines, both SIS and GIS spectra were used;
only the SIS spectra were fit for lower energy lines.
No statistically significant line emission was detected.
The 90\% confidence upper limits on the equivalents widths
of several strong emission lines are listed in Table~\ref{tab:lines}.

We also searched for absorption lines at the redshifted wavelengths
of strong, isolated allowed lines.
The 90\% upper limits on several absorption lines are also listed in
Table~\ref{tab:lines}.

\section{Discussion} \label{sec:discussion}

\subsection{X-ray Properties of the Quasar B2~1028+313}

The total X-ray luminosity of the B2~1028+313/A1030 system was
(0.7,1.3,1.7) $\times 10^{45}$ erg s$^{-1}$ (0.2--10 keV) during
the {\it ASCA}, {\it ROSAT} PSPC, and HRI observations, respectively.
This is in the range of X-ray luminosities for brighter X-ray clusters,
so it was possible that a significant portion of the emission is due
to hot ICM, rather than the AGN.
However, our observations show that the quasar dominates the X-ray
properties of the system, and that X-rays from ICM make at most a
small contribution to the X-ray luminosity.

The X-ray flux of B2~1028+313 varied by a factor of about two
in the one year period between the {\it ASCA} and {\it ROSAT} HRI
observations
(Table~\ref{tab:flux}).
However, the {\it Einstein}, {\it ROSAT} PSPC, and {\it ROSAT} HRI
fluxes are more nearly equal, suggesting that such large variations
may be rare, and that the higher flux (0.7--0.8 $\mu$Jy at 1 keV)
may be more typical.
It is useful to compare this flux density to the core radio flux density,
which is 110 mJy at 5 GHz
(Gower \& Hutchings 1984).
The total flux at this frequency is 161 mJy
(Condon et al.\ 1994), so that B2~1028+313 is core-dominated.
Worrall et al.\ (1994) found a linear correlation between the X-ray and radio
core fluxes of core dominated radio galaxies at redshifts $z > 0.3$.
For the observed core radio flux of B2~1028+313, the
predicted X-ray flux density at 1 keV is 0.01--0.02 $\mu$Jy, which is
a factor of at least 30 smaller than the observed flux.
On the other hand, the observed X-ray flux is similar to that of
lobe-dominated radio quasars with the same radio flux.
This may argue that quasars and radio galaxies have a significant component
to their X-ray emission which is compact and nonthermal, but not
strongly beamed.

The X-ray spectra of B2~1028+313 are generally well-fit by a power-law with
a photon number spectra index of $\Gamma \approx 1.8$, plus an
excess soft X-ray component.
The existing X-ray spectra do not provide a clear determination of the
spectral shape of the soft component, but it can be fit by a blackbody with
a temperature of about 30 eV.
It contributes only about 15\% of the total 0.2-10 keV X-ray luminosity
of the quasar, but provides about 71\% of the flux density at 0.2 keV.

\begin{figure*}[htb]
\vskip4truein
\includegraphics{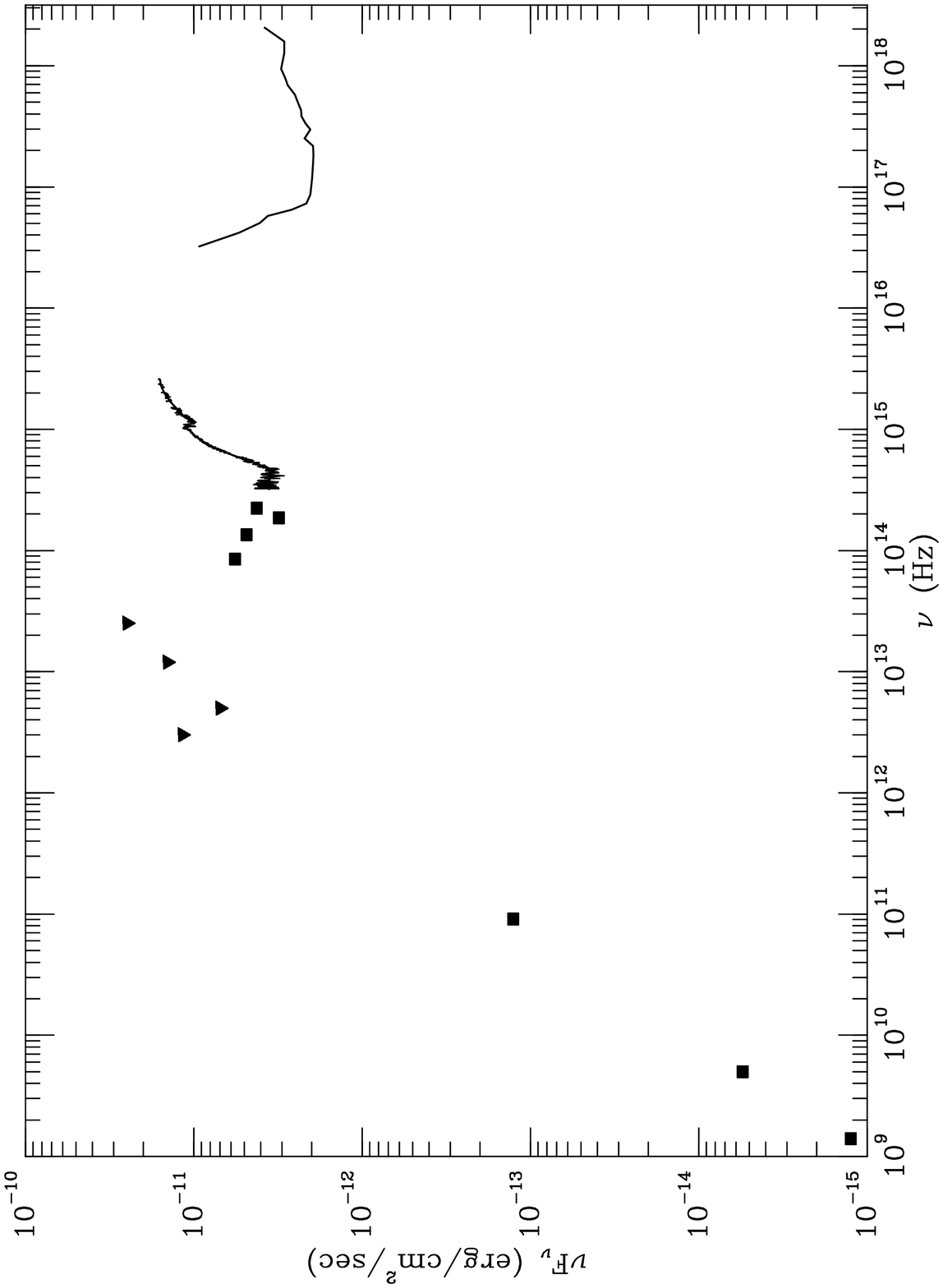}
\caption{
The spectral energy distribution of B2~1028+313.
The frequency times the flux density, $\nu F_\nu$ is plotted versus
the frequency as observed.
The X-ray flux densities are from the unfiltered {\it ROSAT} PSPC data
and the {\it ASCA} data, both corrected for Galactic absorption and
instrumental sensitivity.
The optical and UV data are taken from the composite
{\it HST} and ground-based spectrum in Koekemoer et al.\ (1998).
Emission and absorption line free regions were used to determine this
spectrum, which is corrected for Galactic extinction.
The near-IR, far-IR, and radio core fluxes are from the compilation of
Elvis et al.\ (1994).
The far-IR values, which are shown as inverted triangles,
are upper limits.
\label{fig:sed}}
\end{figure*}

In Figure~\ref{fig:sed}, we show the spectral energy distribution of
B2~1028+313 from the radio through the X-ray spectral band.
We do not show the error bars on the points, because they are generally
small and are difficult to see on the Figure.
Note that the different spectral regions were observed at different epochs,
and variability may affect the comparison.
The X-ray flux densities are from the unfiltered {\it ROSAT} PSPC data
and the {\it ASCA} data presented in this paper.
Because the {\it ASCA} fluxes are lower than the {\it Einstein} or
{\it ROSAT} fluxes (Table~\ref{tab:flux}), we increased the {\it ASCA}
spectrum by a constant factor so that it agreed with the {\it ROSAT}
PSPC spectrum at 1 keV.
The X-ray fluxes were corrected for the response of the two instruments,
and for Galactic absorption.
The {\it ASCA} and {\it ROSAT} spectra were binned to reduce the errors,
and averaged in the common region of spectral coverage.

The optical and UV data are taken from Koekemoer et al.\ (1998), and
are based on {\it HST} and ground-based spectra.
In order to get a measure of the continuum emission from the quasar,
the flux densities were determined in wavelength regions which were free
of strong emission or absorption lines.
The spectrum was corrected for Galactic extinction, assuming
$A_V = 0.104$ mag,
which is appropriate for the Galactic column of
$N_H = 1.98 \times 10^{20}$ cm$^{-2}$.
When this Galactic extinction correction is made,
a broad 2200 \AA\ absorption feature in the UV spectrum largely disappears.
There is a residual feature at slightly longer wavelengths, which might
be due to intrinsic absorption at the quasar redshift, additional Galactic
absorption, or a very broad emission line at longer wavelengths which was
not subtracted from the spectrum.

The near-IR, far-IR, and radio core fluxes are from the compilation of
Elvis et al.\ (1994).
The far-IR values, which are shown as inverted triangles in
Figure~\ref{fig:sed}, are upper limits.
The near-IR values are fairly uncertain because of measurement errors
and a significant correction for the background galaxy emission in
these bands
(see the discussion in Elvis et al.\ [1994]).

The optical and UV measurement also show a large peak in the mid-UV.
This peak is crudely similar to a $\sim$20 eV blackbody, although the
observed peak is considerably broader than a single temperature blackbody.
The spectral index suggested by the near-IR data is considerably steeper
than that for the hard X-ray data.
This may indicate that the non-thermal spectrum flattens with increasing
frequency.
Alternatively, there may be a second peak in the IR spectral band.
However, the uncertainties in the near-IR make it difficult to be certain
about the spectrum there.
The hard X-ray power-law spectrum would extrapolate to flux consistent with
the highest frequency radio measurement, but it is not clear whether the
radio spectrum has turned over by this point.

\subsection{Intracluster Medium in A1030} \label{sec:discussion_icm}

One of the aims of this work was to determine the properties of the
intracluster medium in A1030 by separating the X-ray emission of that
medium from that of the quasar B2~1028+313.
We had hoped to detect intracluster emission spatially (as an extended
component) and spectrally (as a thermal component in the X-ray spectrum).
We did detect an extended component to the X-ray emission, which
was elongated to the NNW and SSW
(Figures~\ref{fig:hri_contour} and \ref{fig:pspc_contour}).
However, there are several features of this emission which are
inconsistent with normal ICM emission from clusters.
First, it is rather highly elongated
(Figure~\ref{fig:pspc_contour}).
Second, its radial surface brightness profile is not well-fit by
either a beta-model or a cooling flow model.
Third, the extended emission is quite soft, and would imply an ICM gas
temperature of $kT \approx 0.2$ keV, which is much lower than typical
temperatures of $kT \approx 7$ keV.

If we exclude the elongated X-ray features in Figure~\ref{fig:pspc_contour},
less than 4\% of the X-ray emission in the {\it ROSAT} band originates
in an extended component beyond a radius of 1\arcmin.
Of course, the brightest cluster X-ray emission may be blended into
quasar emission at the center of the cluster.
If A1030 had ICM with a distribution which followed a beta model
(eq.~\ref{eq:beta_model}) with $\beta = 2/3$ and $r_c > 150$ kpc,
then $>$54\% of the ICM X-ray emission should lie at projected
radii greater than 1\arcmin.
This suggests that $\la$8\% of the total X-ray emission is due
to the ICM, and that an upper limit on the ICM X-ray luminosity is
$L_X (ICM) \la 1.0 \times 10^{44}$ erg s$^{-1}$ (0.2--10 keV).
For a cluster temperature of 7 keV, the corresponding limit on
the central electron density, assuming the same beta model,
is $n_e \la 3 \times 10^{-3}$ cm$^{-3}$.
The limit on the pressure of the ICM is $P_{ICM} \la 7 \times 10^{-11}$
dyne cm$^{-2}$.
If we adopt a lower temperature of 4 keV, which might be more appropriate
given the low X-ray luminosity of the cluster, the limit on the
central electron density is not affected, the limit on the X-ray luminosity
is 20\% lower, and the limit on the central pressure is
$P_{ICM} \la 4 \times 10^{-11}$ dyne cm$^{-2}$.

The X-ray spectra do not require any ICM thermal component.
The upper limits on the X-ray luminosity of a thermal component with
a heavy element abundance of 0.3 of solar and a typical cluster temperature
of $kT = 7$ keV is $L_X < 1.5 \times 10^{44}$ erg s$^{-1}$.
In addition to limits based on overall fits to the spectrum, we also
searched for strong iron, silicon, and sulfur emission lines from the ICM.
None were found;
the limits on the equivalent width are consistent with but somewhat
weaker than the limits on X-ray luminosity based on the integrated
spectrum, assuming typical cluster abundances.
We also placed limits on the cooling rate of any central cooling flow
component.
The upper limit, making the same assumptions about abundances and the
initial temperature, is
$\dot{M} < 93 \, M_\odot$ yr$^{-1}$.

These limits on the ICM X-ray luminosity do not require that A1030 be
unusually faint for an Abell richness 0 cluster.
The upper limit is at the high end of the X-ray luminosities observed
for richness 0 clusters at redshifts below 0.25
(e.g., Rosati et al.\ 1998).

\centerline{\null}

\subsection{Extended X-ray Emission and Radio Source}
\label{sec:discussion_radio}

The elongation of the extended X-ray emission in A1030 is along the
same direction as the extended radio emission.
On small scales ($r \la 5\arcsec$).
the radio image of B2~1028+313 shows a bright core and a jet propagating
to the north at a position angle of about $-30^\circ$
(Gower \& Hutchings 1984).
On larger scales, radio lobes are seen to the NNW and SSE.
While the orientation varies somewhat with radius between positions angles
of about $-45^\circ$ and 0$^\circ$ to the north, and about 130$^\circ$ and
170$^\circ$ to the south, overall the orientation is along position angles
of about $-30^\circ$ and $150^\circ$
(Owen, White, \& Ge 1993;
Owen \& Ledlow 1997).
The largest scale radio emission is most useful for comparison to the
X-ray emission.
On scales comparable to the size of the extended X-ray emission in
the {\it ROSAT} PSPC image
(Figure~\ref{fig:pspc_contour}), the best radio image may be provided by
the NVSS survey
(Condon et al.\ 1998).

\begin{figure*}[tbh]
\vskip4.2truein
\includegraphics{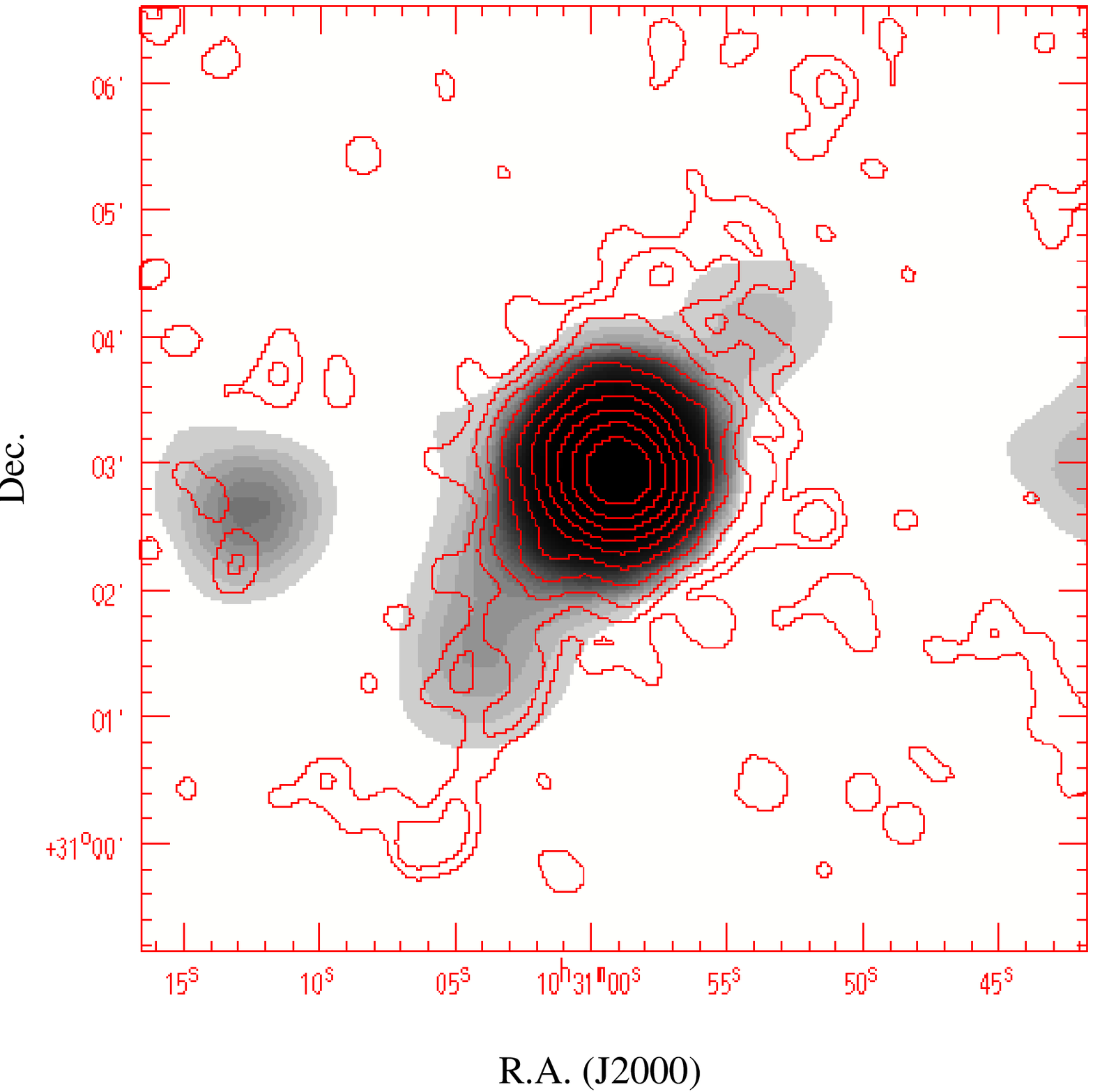}
\caption{
Contours of the central {\it ROSAT} PSPC X-ray image of A1030 are shown
superposed on a greyscale representation of the NVSS 20 cm radio image.
The greyscale for the radio ranges from about 1 to 205 mJy/beam.
The {\it ROSAT} image and contours levels are the same as those in
Figure~\ref{fig:pspc_contour}.
\label{fig:nvss_pspc}}
\end{figure*}

Figure~\ref{fig:nvss_pspc} shows contours of the {\it ROSAT} image superposed
on a greyscale representation of the NVSS 20~cm radio images.
The NVSS image has a resolution of 45\arcsec.
The radio source to the east is almost certainly an unrelated background
source.
At radii of 1--2\arcmin, the radio emission is elongated in nearly
the same direction as the X-ray emission.
It is not clear whether the radio emission continues to larger radii;
if one extrapolates the observed radial surface brightness trend to larger
radii, the emission might be expected to be below the noise in the NVSS
image.

The total flux density of the radio source is 254 mJy at 20 cm
(Owen \& Ledlow 1997).
In the NVSS image, 215 mJy of emission is due to unresolved emission
from the central source, leaving a total of 39 mJy of possible extended
emission on the scale of the 45\arcsec\ beam of the survey.
The fluxes in the northern and southern extensions on the NVSS image
in Figure~\ref{fig:nvss_pspc} are about 5.4 and 7.4 mJy, respectively.

Extended emission is also seen at lower frequencies in the
49 cm WENSS survey image of this region
(Rengelink et al.\ 1997).
Although the noise level is higher and the beam larger than in the
NVSS survey, the WENSS image shows north and south lobes at
the same locations as in the NVSS image.
The flux densities of the northern and southern lobes at 49 cm
are 22.5 and 20.3 mJy, respectively.
Because of differences in the resolution, it is unclear whether the NVSS and
WENSS fluxes correspond to identical regions.
To reduce somewhat the uncertainties, we will sum the emission from
the two lobes in each of the instruments.
Assuming they come from the same region, the ratio of fluxes implies
an energy spectral index in the radio of $\alpha = 1.45 \pm 0.30$
(a photon number spectral index of $\Gamma = 2.45 \pm 0.30$), where
the error estimate follows from the different values for the northern and
southern lobes.
This is similar to, but not quite as steep as, the spectral index suggested
by the X-ray colors of the extended emission
(\S~\ref{sec:spatial_pspc}).
Given the fluxes, size, and spectral index, we can estimate the properties
of the radio source under the minimum energy assumption
(e.g., Miley 1980).
We assumed that the relativistic
electrons and ions had equal energies, that the radio plasma filled
the observed emitting volume, and that the magnetic field was
transverse to the line of sight.
The radio spectrum of each component was assumed to be
a power law from a frequency of 10 MHz to 100 GHz.
We assumed that the observed lobes were cylinders whose axis
is perpendicular to our line of sight.
The implied magnetic field strength was $B \approx 1.6$ $\mu$G,
the nonthermal pressure was $P_{rad} \approx 1.5 \times 10^{-13}$
dyne cm$^{-2}$, and the total energy in the lobes was about
$1.6 \times 10^{59}$ erg.
The lifetime of the electrons which emit at 1.4 GHz is about
$4 \times 10^7$ yr, which is probably an upper limit to the age of
the radio source.

What is the source of the elongated X-ray emission which is extended
along the radio structure in Figure~\ref{fig:nvss_pspc}?
One possibility is that the radio lobes are interacting with
intracluster gas, and have compressed the gas.
In this case, the enhanced X-ray emission would be expected around
the perimeter of the radio lobes.
The X-ray and radio spatial resolutions in Figure~\ref{fig:nvss_pspc} are too
poor to provide a useful test of this predictions.
If the radio lobes compressed the surrounding intracluster gas (either
adiabatically or in shocks), the gas would have been heated.
Actually, the X-ray emission from the X-ray extensions is quite soft
(\S~\ref{sec:spatial_pspc}), corresponding to a gas temperature of
about 0.2 keV.
Thus, this model requires that the compressed gas cool to low
temperatures.
No evidence of cooling emission is seen in the spectrum of the
source (\S~\ref{sec:spectra_both}).
The compressed gas would have to be dense enough to cool within the
lifetime of the radio source, which would require that
$n_e \ga 0.1$ cm$^{-3}$.
If the intracluster gas were compressed by strong shocks, its density
would only increase by a factor of 4 until cooling was effective.
Thus, the ambient gas density in the region around the radio source
must exceed $2 \times 10^{-2}$ cm$^{-3}$ for this model to work.
The upper limit on the gas density at the radii of the radio lobes
based on the upper limit of the X-ray surface brightness at these radii
away from the radio source is $n_e \la 1 \times 10^{-3}$ cm$^{-3}$.
Thus, it is unlikely that such gas could cool rapidly enough to
produce the regions of enhanced soft X-ray emission.

Alternatively, the X-ray emission may be due to cooler gas which is falling
into the central galaxy in A1030, either alone or as part of the merger
of a gas-rich galaxy with the central cD.
West (1994) suggested that such mergers onto cD galaxies will mainly occur
along filaments in the large-scale structure surrounding the cluster,
and that they will produce cD galaxies which are prolate in shape.
West argued that gas which falls into such a galaxy will form a disk
with its rotation axis being parallel to the major axis of the cD galaxy.
The extended X-ray emission in A1030 is elongated at an angle of
about 70$^\circ$ from the major axis of the host galaxy of B2~1028+313,
which is at a position angle of 45$^\circ$
(Ledlow \& Owen 1995).
Thus, this might be consistent with West's scenario.
However, West also argues that the gaseous accretion disk will determine
the axis of the central radio jets, which should be perpendicular to the
major axis of the gas disk and parallel to the optical major axis of
the galaxy.
This gives the wrong relative orientations of the gas disk and radio
source and of the optical galaxy and radio source for B2~1028+313.

Finally, the extended X-ray emission might be due to Inverse Compton
(IC) scattering of Cosmic Microwave Background (CMB) photons by low energy
cosmic ray electrons from the radio lobes.
As noted above (\S~\ref{sec:spatial_pspc}),
the extended emission has a very soft X-ray color, which is consistent
with a steep power-law X-ray spectrum ($\Gamma = 2.9 \pm 0.4$).
We have calculated the expected ratio of the IC X-ray flux to the
synchrotron radio fluxes using the general expressions for the
IC and synchrotron emission from power-law electron energy distributions
(Blumenthal \& Gould 1970;
Sarazin 1988).
The ratio of the IC flux at a frequency $\nu_X$ to the synchrotron flux
at a frequency $\nu_r$ is
\begin{eqnarray}
\frac{f_{IC} ( \nu_X )}{f_{syn} ( \nu_r )}  & = &
\frac{2 \pi e}{h^2 c^4 m_e B_\perp} \,
\frac{b(p)}{a(p)} \,
\left( k T_{CMB} \right)^3 \,
\left( 1 + z \right)^{3 + \alpha} \nonumber \\
& & \qquad \mbox{} \times
\left( \frac{ 4 \pi m_e c k T_{CMB}}{3 h e B_\perp} \right)^\alpha
\left( \frac{ \nu_r }{ \nu_X } \right)^\alpha \, ,
\label{eq:fluxratio}
\end{eqnarray}
where $T_{CMB} = 2.73$ K is the temperature of the CMB radiation
at zero redshift, $B_\perp$ is the component of
the magnetic field in the radio source perpendicular to our line of sight,
$p = 2 \alpha + 1$ is the power-law index of the electron energy distribution, 
and $a(p)$ and $b(p)$ are two analytic functions of $p$, which are given
in equations~(5.8) and (5.6) in Sarazin (1988).
The power-law electron energy distribution was determined from the
radio flux and spectral index $\alpha = 1.45$ of the radio lobes,
assuming that the electrons were fit by a single power-law distribution.
Given the radio fluxes of the two lobes, the radio spectral index,
and assuming a magnetic field of $B_\perp = 1 \, \mu$G, the total predicted IC
flux density from both of the lobes is 0.3 $\mu$Jy at 0.2 keV,
which is about a factor of two smaller than the observed total flux
of both lobes.
Because of the uncertainties in all these numbers, it is possible that the
extended X-ray emission may be inverse Compton emission associated with
the radio source.
The IC soft X-ray emission might also be enhanced by scattering of the radio
emission from the quasar, particularly if part of that emission is beamed in
the same direction as the radio beams.
If the spectral energy distribution of the quasar (Figure~\ref{fig:sed})
is extrapolated to the peak of the CMB spectrum, one estimates that
the brightness temperature of the radio source is about 1 K
at a distance corresponding to a projected radius of 1\arcmin\ 
from the nucleus.
Thus, the local radiation field probably does increase the IC flux somewhat.
One concern with the IC model is that one might expect a more exact
correspondence between the extended X-ray and radio features in
Figure~\ref{fig:nvss_pspc}.
On the other hand, the differences might be due to the low spatial
resolution and signal to noise ratio of both the X-ray and radio images, or
to differences between the distributions of the low energy cosmic
rays which produce the IC X-rays and the higher energy particles which
produce the synchrotron radio emission.
Another concern with the IC model is that it requires that the same
very steep spectral index for the radio lobes apply down to very low
frequencies, since the electrons which produce the IC soft X-ray emission
have much lower energies than those which produce the radio emission.

\acknowledgments

We thank Dan Harris, Jonathon Silverman, and the referee, Patrick Hall, for
useful comments.
C. L. S. was supported in part by NASA ROSAT grants
NAG 5--3308,
NAG 5--4787,
NASA ASCA grant NAG 5-2526,
and NASA Astrophysical Theory Program grant 5-3057.
Partial support for this work was provided by NASA through grant number 
GO-05934.01-94A from the STScI, which is operated by AURA, under NASA 
contract NAS5-26555.

\begin{figure*}[tbh]
\vskip3.5truein
\includegraphics{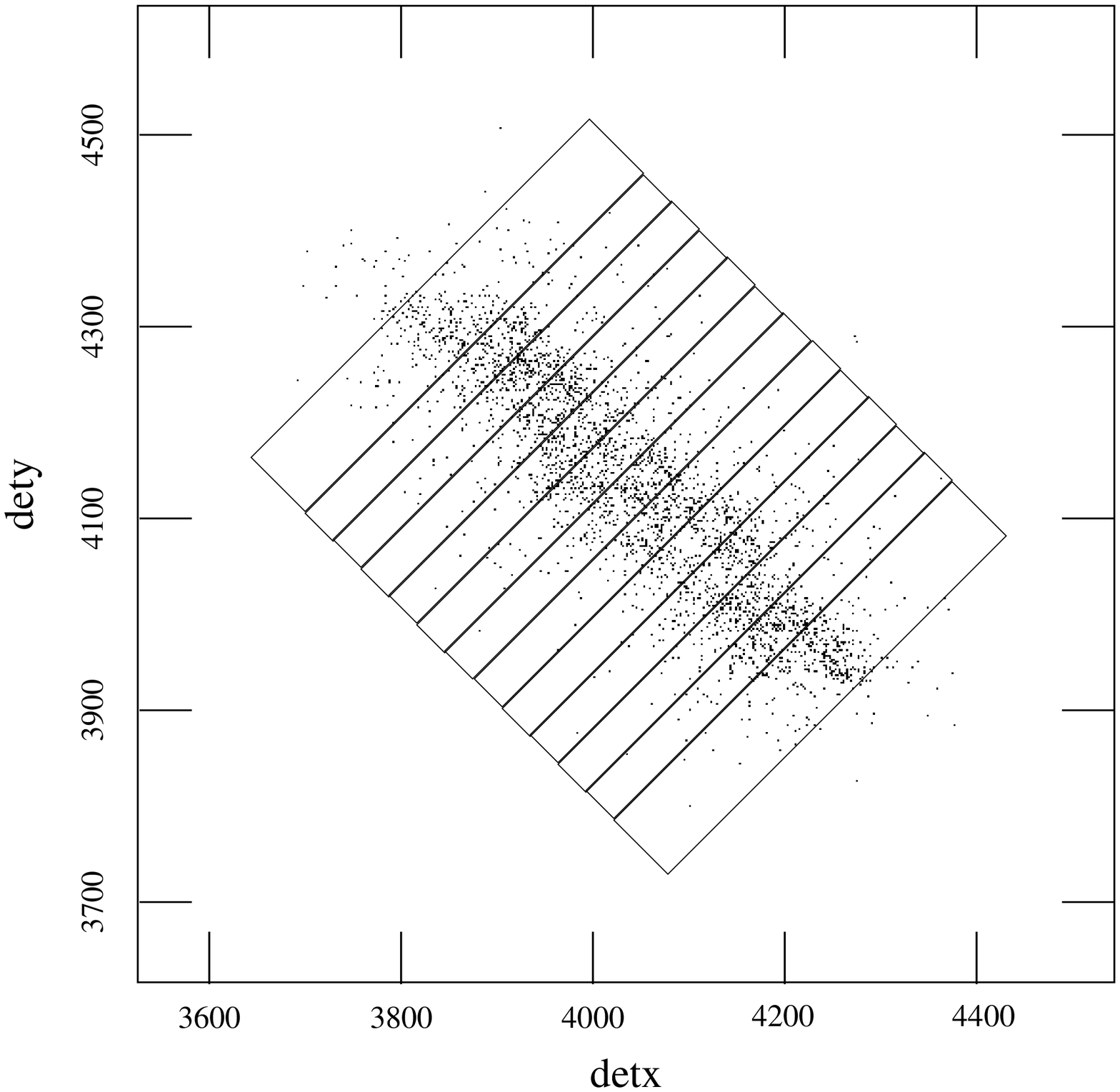}
\caption{
The image of photons from the center of A1030 in HRI detector coordinates.
The elongation of the image at a position angle of about $45^\circ$ is
due to the ``wobbling'' of the telescope.
The rectangles show the regions of the detector used to collected subimages
at different phases of the wobble.
\label{fig:wobble}}
\end{figure*}

\appendix

\section{Correction of HRI Image for Aspect Errors} \label{sec:aspect}

During pointed observations with {\it ROSAT}, the satellite is
normally ``wobbled'' by several arcmin with a period of several
hundred seconds.
The result of this wobbling is  that photons from a given location on
the sky are collected over an elongated regions of the detector.
Figure~\ref{fig:wobble} shows the detector coordinate positions of
photons from the center of the image of A1030.
The elongation at $PA \approx 45^\circ$ is due to the wobble.
The image is reconstructed by determining the pointing direction of
the satellite at the arrival time of each of the photons.
Morse (1994) suggested that errors in this aspect solution during the
wobble can result in extended and elongated images of point sources.
(Note that Morse finds that the observed elongations
of the images of point sources aren't necessarily in the same direction
as the projection of the wobble on the sky, although the elongations are still
due to wobble-dependent aspect errors.)

Morse argued that aspect errors were correlated with the phase of the
wobble, and they might be corrected by registering subimages collected
as a function of the wobble phase.
He suggested that subimages be collected from small regions of the
detector.
Because the extension  of a point source in detector coordinates is
mainly due to the wobble, this is nearly equivalent to selecting
photons at the same phase in the wobble.
Morse suggested that regions which were $5\arcsec \times 5\arcsec$ be used.
The subimages must contain at least 100 photons to allow the
centroid of each to be determined with sufficient accuracy.
Even if larger detector regions of $10\arcsec \times 10\arcsec$ are used,
there will typically be $\ga 100$ such subimages.
Thus, this technique requires that the central source in the
image contain $\ga 10^5$ photons.
We attempted to correct our image for A1030 using the FTOOLS
HRIASPCOR implementation of Morse's algorithm.
Unfortunately, the central region of our image of A1030 contained
about 3000 counts, including background.
We found that this was too few counts to apply Morse's algorithm
in its original form.
The HRIASPCOR program produced a highly elongated and irregular image.

We developed a minor variant of the Morse algorithm, which can be applied
to somewhat fainter sources.
(Recently, we found that a similar technique has been used previously by
Hall et al.\ [1995].
Also, as this paper was being revised, we learned of a new technique for
improving the aspect solution of HRI images, which is presented in
Harris, Silverman, \& Hasinger [1998].
In principle, this new technique should work even better than the previous
ones, including the one presented here, for relatively faint sources.)
Because the wobble is mainly in a single direction, we collect photons
into subimages within the detector using long, narrow rectangle regions.
The longer side of these rectangular regions is perpendicular to the
direction of the wobble.
Figure~\ref{fig:wobble} shows the 13 rectangular regions we used to
correct the image of A1030.
The regions are 20\arcsec, except for the two at the ends of the wobble
which are 40\arcsec\ wide.
We constructed subimages using photons collected from each of these
detector regions, determined their centroids, and registered them
to a common centroid.
The resulting image of the center of A1030 is shown in
Figure~\ref{fig:hri_contour}b.
The center part of this image is narrower than
the original image (Fig.~\ref{fig:hri_contour}a).

One concern with the Morse procedure is that registering subimages
of an image will always produce a more compact image, even if the
subimages are randomly sampled from the original image.
As an extreme example, if the number of subimages equals the number
of photons, then the resulting corrected image will be a delta function,
whatever the distribution of the original photons.
In general, if an image contains $N$ photons with a gaussian distribution,
and the image is subsampled randomly into $m$ subimages each containing
$(N/m)$ photons, then the dispersion of the gaussian distribution will
be artificially reduced by a factor of $[(N-m)/(N-1)]^{1/2}$.
In our case, $N \approx 3000$ and $m = 13$, so this should not be
an important effect.
We have verified this by creating random subimages of our image,
and we find that they are not significantly different from the original
image, and are not as centrally condensed as the wobble-corrected
image in Figure~\ref{fig:hri_contour}b.
A second worry is that the distribution of photons on the detector
is a convolution of the wobble and the actual spatial distribution of
photons on the sky.
As a result, registering the subimages artificially reduces the extent
of the resulting image.
However, this effect depends on the ratio of the size of the image
on the sky ($\sim$10\arcsec\ for the central part of the A1030 image)
to the size of the wobble ($\sim$5\arcmin\ in Figure~\ref{fig:wobble}),
and thus is expected to be quite small.

\centerline{\null}
\centerline{\null}
\centerline{\null}

\end{document}